\documentclass[12pt, openright, mydesign]{article}

\usepackage{etoolbox}
\newbool{ISTHESIS}
\boolfalse{ISTHESIS} 
\usepackage{etoolbox}

\ifbool{ISTHESIS}{
  \newcommand{\toplevel}{\chapter}
  \newcommand{\secondlevel}{\section}
  \newcommand{\thirdlevel}{\subsection}

}{
  \newcommand{\toplevel}{\section}
  \newcommand{\secondlevel}{\subsection}
  \newcommand{\thirdlevel}{\subsubsection}

}

\usepackage[utf8]{inputenc}
\usepackage[T1]{fontenc}
\usepackage{lmodern}
\usepackage{mwe}

\usepackage{geometry}
\usepackage{forest}
\forestset{
  folder/.style = {font=\ttfamily\bfseries}, 
  file/.style   = {
    font=\ttfamily,
    align=left,
    text width=0.82\linewidth, 
    inner xsep=0pt
  }
}

\usepackage[dvipsnames]{xcolor}
\usepackage{soul}
\usepackage{todonotes}

\usepackage{listings}
\usepackage{inconsolata} 
\usepackage{upquote}     

\definecolor{codebg}{gray}{0.97}
\definecolor{codenumbers}{gray}{0.45}
\definecolor{codestring}{rgb}{0.58,0,0.82}

\lstdefinestyle{thesis}{
  language=Python,
  basicstyle=\ttfamily\footnotesize,
  numbers=left,
  numberstyle=\scriptsize\color{codenumbers},
  numbersep=10pt,
  frame=single,
  framerule=0.2pt,
  backgroundcolor=\color{codebg},
  showstringspaces=false,
  keepspaces=true,
  columns=fullflexible,
  tabsize=2,
  breaklines=true,
  breakatwhitespace=false,
  captionpos=b,
  keywordstyle=\bfseries,
  commentstyle=\itshape\color{codenumbers},
  stringstyle=\color{codestring},
  upquote=true,
  float,
  floatplacement=tbp,
}
\usepackage{setspace}
\usepackage{andersen-math}

\usepackage{amsmath, amssymb}

\usepackage{booktabs}
\usepackage{tabularx}
\usepackage{array}
\newcolumntype{Y}{>{\raggedright\arraybackslash}X}
\newcolumntype{T}{>{\ttfamily\footnotesize\raggedright\arraybackslash}X}

\usepackage{graphicx}
\usepackage{subcaption}

\usepackage{xurl}
\usepackage{hyperref}
\hypersetup{
  pdftitle={The Curious Case of the Default Settings},
  pdfauthor={Madelyn Victoria Andersen},
  pdfsubject={Variational Inference},
  pdfkeywords={Massachusetts Institute of Technology, MIT},
  colorlinks=true,
  linkcolor=MidnightBlue,
  citecolor=MidnightBlue,
  urlcolor=MidnightBlue
}

\usepackage[round,authoryear]{natbib}
\usepackage[nameinlink,capitalise,noabbrev]{cleveref}

\newtheorem{innercustomgeneric}{\customgenericname}
\providecommand{\customgenericname}{}
\newcommand{\newcustomtheorem}[2]{%
  \crefname{#2}{#2}{#2s}%
  \newenvironment{#1}[1]
  {%
   \renewcommand\customgenericname{#2}%
   \crefalias{innercustomgeneric}{#2}%
   \renewcommand\theinnercustomgeneric{##1}%
   \innercustomgeneric
  }
  {\endinnercustomgeneric}
}
\newcustomtheorem{customdefinition}{Definition}
\newcustomtheorem{customlemma}{Lemma}
\newcustomtheorem{customproposition}{Proposition}

\makeatletter
\@ifundefined{CopyrightYear}{\newcommand{\CopyrightYear}[1]{}}{}
\makeatother

\usepackage[T1]{fontenc}
\usepackage{lmodern}
\usepackage[final]{microtype}

\usepackage{amssymb}

\usepackage[round,authoryear]{natbib}
\usepackage{hyperref}
\usepackage[nameinlink,capitalise,noabbrev]{cleveref}

\usepackage{graphicx}
\graphicspath{{figs/}{figures/}{images/}{convergences/}}
\usepackage{subcaption}

\usepackage{booktabs}
\usepackage{tabularx}
\usepackage{array}
\newcolumntype{Y}{>{\raggedright\arraybackslash}X}
\newcolumntype{T}{>{\raggedright\arraybackslash}p{0.30\linewidth}}

\usepackage{caption}
\usepackage{enumitem}
\setlist{nosep,leftmargin=2em}

\usepackage{siunitx}
\usepackage[dvipsnames]{xcolor}

\newcommand{\ELBO}{ELBO}
\newcommand{\ELBOspace}{ELBO }
\newcommand{\KL}{\ensuremath{\mathrm{KL}}}

\newcommand{\underlyingmu}{\text{$\tilde{\mu}$}}
\newcommand{\underlyingsigma}{\text{$\tilde{\sigma}$}}
\newcommand{\mcgradnodes}{Monte Carlo gradient samples }

\newcommand{\mcnode}{MC sample }

\newcommand{\mcnodes}{MC samples }

\newcommand{\mcnodenospace}{MC sample}

\usepackage{cleveref}

\usepackage{placeins}

\begin{document}

\title{The Curious Case of the Default Settings: Evaluating Default Performance of Variational Inference
Software}

\author{Madelyn Andersen, Elizabeth Bersson, Tamara Broderick\\
MIT LIDS}

\CopyrightYear{2025} 

\date{}

\maketitle

\begin{abstract}
We systematically evaluate a suite of off-the-shelf variational inference (VI) software packages from the perspective of a standard practitioner. Using simple analytic benchmark models, we assess the accuracy and stability of the default VI settings in PyMC, NumPyro, and TensorFlow Probability. Unlike previous research focusing on methodological advances, our evaluation emphasizes software implementation and the default configurations that typical users encounter. Our results show that default settings can yield biased approximations of posterior summaries even for simple one-dimensional conjugate models, controls of initialization and transformations differ between software implementations, and relying on defaults may yield silent failures or poor approximations.
\end{abstract}


\toplevel{Introduction}
\label{sec:intro}
Bayesian methods underpin decision-making in healthcare \citep{blaiotta2016variational}, climate science \citep{palma2025data}, and financial risk assessment \citep{arian2020encoded}.  In principle, full Bayesian inference provides access to the exact posterior distribution over model parameters, enabling the computation of any posterior summary of interest, including posterior means, variances, credible intervals, and predictive probabilities \citep[Ch.~1.1, Ch.~4~intro paragraph]{hoff2009firstch4}. In applications, posterior means and variances are commonly reported summaries of Bayesian inference; this practice is well-established in the literature and often guides interpretation and downstream decisions (\citealp[Ch.~7]{SASIntroBayes2013};
 \citealp[Sec.~10.1.5]{RossBayesianReasoning2022};
 \citealp[Sec.~3.7]{Mason2018BayesianFramework}). 

Closed-form solutions, however, are rarely available for posterior distributions, or for posterior means and variances, in practically relevant models \citep[Ch.~4]{hoff2009firstch4}, so practitioners often turn to approximations. One such option is Markov Chain Monte Carlo (MCMC). 
MCMC approximates exact posterior expectations by generating samples \citep[Sec.~1]{Roberts2004general} and is asymptotically exact as the number of samples increases (\citealp{nemeth2019stochastic}; \citealp[Sec.~3.2]{Roberts2004general}). In practice, however, the computational cost is often prohibitive for complex models and large datasets, as in many real-world applications \citep{Giordano2024BlackBox, salimans2015markovchainmontecarlo}.

In complex or large-data cases where MCMC struggles, analysts often to turn to variational inference (VI) in the hope of using fewer computational resources than sampling methods \citep{kucukelbir2016automatic, ranganath2014black}. 
Variational inference replaces sampling with an optimization problem: choose a variational family and minimize a divergence to the posterior over this family \citep{jaakkola1997variational, jordan1999introduction}. In practice, the reverse Kullback--Leibler (\KL) divergence is commonly chosen \citep[see][ch.~10]{bishop2007ch10}. When the family is restrictive, the approximation for a posterior mean or variance can be systematically biased \citetext{see  \citealp{bishop2007ch10} ch.~10}. While VI is often faster than MCMC \citep{salimans2015markovchainmontecarlo}, the accuracy of the result depends on the choice of variational family and on optimization success, sometimes trading accuracy for greater speed \citep{kucukelbir2016automatic}.

Probabilistic programming languages such as Stan \citep{carpenter2017stan}, PyMC \citep{salvatier2016probabilistic}, NumPyro \citep{bingham2019pyro}, and TensorFlow Probability \citep{abadi2016tensorflow} provide built-in implementations of VI. While VI implementations in many popular packages have made VI methods accessible to non-experts, 
 users often encounter difficulties with the reliability and quality of results, especially when using default settings.  Reports of optimization failures, numerical instability, and poor accuracy in posterior approximations frequently arise in software forums \citep{mcstan_cmdstanpy_variational_density_params_2024, cmdstanpy_dev_2022_vi_example, mcstan32449, mcstan4150}, blog posts \citep{vehtari_2022_advi_dropped}, developer messages \citep{cmdstanpy_dev_2022_vi_example}, and other platforms \citep{stackexchange110369, arvizissue498}. 

 Although systematic evaluations for MCMC software performance exist \citep{thompson2010graphical, kavian2024new}, VI’s distinct optimization-based nature differs from the sampling-based nature of MCMC. It therefore remains an open question how well default settings in current VI software serve users’ needs. The present work evaluates and compares the current default settings available in VI software packages; in particular, we examine optimizer choice, Monte Carlo sample size, control over initialization, control over transformations, and diagnostics for tracking. We assess the quality of posterior approximation under simplified use cases. We find that there is substantial room for improvement in output quality by making straightforward changes to these default settings. In concurrent work, \citet{campbell_placeholder} provide a detailed comparison of a wide range of different optimizers and optimizer settings that might be used within VI (and maximum a posteriori computation); they consider optimizers beyond those currently used as software defaults, and they vary internal settings of the optimizers. They also focus on minimization of VI’s optimization objective whereas we focus on quantities users will commonly report from variational inference (posterior means and variances). Where our focuses overlap (study of current default optimizers in existing software), our recommendations agree; beyond that, our findings are complementary and suggest useful settings for VI software developers.

In the remainder of this paper, we evaluate existing variational inference software packages by assessing their quality of posterior approximation under simplified use cases.  We focus on the default user experience, i.e., the experience under standard settings.  We show that, beyond the variational family, implementation choices (default values, optimizer, and transformations) materially affect accuracy under defaults.

Specifically, in the remainder of this paper, we demonstrate the following aspects of default VI software behavior. These include failure modes that arise in at least one software package (e.g., default settings inducing bias, lack of transformation control).
\begin{itemize}[leftmargin=2em]
  \item \textbf{Default settings can induce bias.} In certain simple models, we find that defaulting to a single Monte Carlo sample and the Adagrad\_Window optimizer yields posterior summary estimates that are systematically biased away from the best variational posterior in certain software implementations.
  \item \textbf{Sensitivity to initialization.} User control over initializations may mitigate posterior summary estimates getting stuck in a (poor) local optimum.
  \item \textbf{Lack of transformation control can cause silent failure.} We find cases where defaulting to no transformations leads to silent VI failure and nonsensical posterior summary values.
  \item \textbf{Some useful diagnostics are missing.} A lack of diagnostics for tracking parameter values in their original space (rather than in an internal representation of the software) may make interpretation more difficult.
\end{itemize}

In \Cref{sec:rel_work}, we review VI, popular existing implementations, and our selection methodology for choosing software packages to evaluate. In \Cref{sec:results}, we compare the implementation differences in each of the selected software packages and present our results from each software package.
We present a concise guide for users in \Cref{sec:prac_guide}, which includes troubleshooting and getting-started tips, concluding with an illustration of some of the limitations of our work in \Cref{sec:conclusion}.
\FloatBarrier

\section{Background and Related Work}\label{sec:rel_work}

Modern probabilistic programming languages (PPLs) aim to make Bayesian inference accessible.
Markov Chain Monte Carlo (MCMC) is regarded as the gold standard for accuracy \citep{nemeth2019stochastic}, but it can be 
prohibitively
computationally expensive \citep{wang2019frequentist}. Variational inference (VI) aims to address this computational challenge by turning inference into optimization. VI fits a tractable approximation to the posterior by maximizing an evidence lower bound (\ELBO). The quality of the approximation depends on the variational family (see \Cref{sub:varfams}), support transformation, and optimization algorithm, choices that differ across software implementations.

\subsection{Variational Inference}
In Bayesian inference, the posterior distribution updates beliefs about latent variables $Z \in \mathbb{R}^d$ after observing data $Y = \{y_1, \ldots, y_n\}$. Consider a probabilistic model in which the latent variables $Z$ are given prior distributions $p(Z)$ and $Y = \{y_1, \ldots, y_n\}$ is a set of observations $y_{i}$ belonging to some measurable space. Each observation $y_i$ is drawn i.i.d.\ from a likelihood $p(y_i|Z)$, and our posterior distribution is $p(Z|Y)$. We will denote the mean and covariance of the posterior distribution as $\mu_p\in\mathbb{R}^d$ and $\Sigma_p\in\mathbb{R}^{d\times d}$, respectively, which are values that practitioners use to make decisions in real-world applications \citep{speagle2020conceptual}.

For many models used in practice, computing the exact posterior mean and covariance is an intractable computation \citep[Ch.~10]{bishop2007ch10}. We may therefore approximate the posterior mean and covariance, possibly using MCMC or VI. MCMC generates samples from the posterior \citep{salimans2015markovchainmontecarlo}. For a specified family $\mathcal{Q}$, VI minimizes the reverse Kullback Leibler (\KL) divergence, $\mathrm{KL}\!\left(q\|p(\,\cdot\,\mid Y)\right)$ across $q\in\mathcal{Q}$, which is equivalent up to a constant to maximizing the Evidence Lower Bound (\ELBO) \citep[Ch.~10, Eqn.~(10.3)]{bishop2007ch10}.

\paragraph{Variational Families}\label{sub:varfams}  
To maximize the \ELBO, we must choose a family  $\mathcal{Q}$ of posterior approximations. A common choice is the mean-field variational family, which partitions latent variables $Z$ into disjoint groups $Z_i$ and factorizes, for a distribution $q\in\mathcal{Q}$,  
\begin{align}
q(Z) = \prod_i q_i(Z_i).
\end{align}  
In its most general form, the mean-field family imposes no functional form restrictions on the $q_i$, but in practice it is often used together with the further restriction that the $q_i$ take an exponential-family form \citep[Sec.~10.1.1]{bishop2007ch10}, \citep[Sec.~2.3]{blei2017variational}.  

For common variational family choices, the \ELBOspace has no closed form. In practice, VI implementations often estimate the gradient of the \ELBOspace numerically at each iteration, commonly with Monte Carlo sampling \citep{ranganath2014black}, and optimize with stochastic gradients \citep{blei2017variational}. 

\paragraph{Best Variational Approximation}
To summarize, the variational inference problem posed by VI is to choose the distribution within a variational family that maximizes the ELBO. In general, this maximizer need not be unique; however, in this paper, we consider cases where it is. Therefore, going forward, we will refer to the ELBO maximizer as the best variational approximation.

Because the exact posterior may fall outside the chosen variational family, the best variational approximation need not equal the exact posterior. Consequently, if a variational approximation returned by software differs from the exact posterior, that difference a priori can be due both to (1) issues inherent in using any variational approximation, even with perfect optimization capability, and (2) issues with trying to solve the variational approximation in practice. Since the first issue has been well-studied in the literature \citep{blei2017variational, turner_sahani_2011_vem}, \citep[Ch.~10.1]{bishop2007ch10} \citep[Ch.~33]{mackay_2003_ita}, we focus on the second issue in the present manuscript. In particular, when evaluating software in what follows, we will treat the best variational approximation as ground truth. In general cases, such ground truth is inaccessible, much like the exact posterior is inaccessible. We will focus on cases where we have access to a unique, analytically tractable solution, enabling us to compare the outputs of VI optimizers directly to this target.

\paragraph{Optimizing in Practice}
Standard optimizers include stochastic gradient descent (SGD, first introduced by \citet{robbins1951stochastic})  and adaptive methods (e.g., Adam, Adagrad, Adadelta, introduced in \citet{kingma2015adam} and \citet{duchi2011adaptive}).  See \citet{campbell_placeholder} for a comprehensive evaluation of optimizers for the purposes of minimizing the variational Bayes objective. Our evaluation focuses on comparing the package defaults, which include Adam and Adagrad\_Window. We note that our evaluation focuses on the quantities users will commonly report from variational inference (namely, posterior means and variances); \citet{campbell_placeholder} instead focus more directly on VI's optimization objective in the form of the ELBO (and the squared gradient 2-norm). \citet{Giordano2024BlackBox} have previously observed that performance of VI methods evaluated in these separate ways can differ.

\paragraph{ADVI in practice} Automatic Differentiation Variational Inference (ADVI) implements a mean-field approximation\footnote{Another form of ADVI allows a full-rank approximation, which PyMC, TFP, and NumPyro all offer as a non-default option.} with normal marginals in a transformation of the original parameter space, then optimizing the \ELBOspace with stochastic gradients \citep{kucukelbir2016automatic}. Essentially, each step of the optimization estimates the gradient with a pre-specified number of Monte Carlo samples.

Model parameters may have constrained supports (e.g., variances restricted to $\mathbb{R}_{>0}$, simplex-valued probabilities, or correlation matrices), so ADVI applies a bijective transformation only to those parameters with constrained domains, mapping them into an unconstrained Euclidean space. Concretely, ADVI samples
\begin{align}\label{eq:advi}
z_i^\ast \sim \mathcal{N}(\underlyingmu, \underlyingsigma^2), \qquad
z_i = T(z_i^\ast),
\end{align}
where $z_i^\ast \in \mathbb{R}^{d^\ast}$  is drawn from a normal distribution with mean $\tilde{\mu}$ and standard deviation $\tilde{\sigma}$, and $T$ maps each unconstrained parameter to the constrained space.   Optimization occurs in the unconstrained space, while practitioners often interpret posterior summaries in the constrained domain. When we optimize, we seek stabilization, which we roughly define as the point at which successive changes in the object of interest become negligible after a finite number of iterations; we hope this stabilization indicates that the optimization algorithm has at least found a local optimum, and ideally a global optimum.

Transformations are essential because many real-world models impose constraints (e.g., heights are strictly positive, blood pressure values lie within an interval, species proportion vectors must lie in the probability simplex). Typical bijectors include exponential for positivity, $\mathrm{sigmoid}$ for the unit interval, stick-breaking for simplices, and Cholesky-based maps for covariance matrices \citep{stan2025usersguide, dillon2017tensorflow, phan2019composable}.

VI has several well-known limitations. 
For example, the nonconvex, mode-seeking nature of the reverse-\KL\ can favor only a few high-density modes instead of the full posterior \citep{zhang2019advances}. Further, VI is well-known to underestimate marginal variances \citep{giordano2018covariances, mackay_2003_ita, giordano2016fast, wang2005inadequacy, giordano2015linear}\citep[Ch.~10]{bishop2007ch10}, and increasing expressiveness of the variational family may not lead to better posterior estimates \citep{turner_sahani_2011_vem}. When stochastic gradients are used with ADVI, the resulting posterior variances and ``convergence'' behavior may be unreliable or unclear \citep{Giordano2024BlackBox}. We do not elaborate on these fundamental limitations of VI here. Rather, they motivate our study: since VI can be sensitive to model structure and optimization pathologies, we would like to understand how different software implementations perform in practice when confronted with such challenges.

\subsection{Evaluating Approximation Quality of Implementation Choices}
We evaluate PyMC, NumPyro, and TensorFlow Probability (TFP), three widely used Python probabilistic programming frameworks (PPLs) with active support for VI.  
Though we exclude Stan’s VI module in this draft, we plan to evaluate it in our future work.

Although other PPLs (e.g., Bambi, Edward) are part of the broader ecosystem \citep{strumbelj2024past}, we restrict scope to the three packages above because---apart from Stan---they represent the most popular subset of software packages including unique VI implementations as of October~2025, according to the PyPI statistics database and CRAN logs (\Cref{table:downloads}) \citep{pypistats_zenodo, cranlogs}. 
We cannot directly compare the download numbers reported for Python to those reported for R, since what constitutes a ``download'' may differ by source \citep{PyPA_PyPI_downloads}. We note that all three software packages we consider offer functionality beyond VI (e.g., MCMC) and many users may have downloaded them for a use other than VI. Practitioners primarily use either Python or R for variational inference; we have chosen to limit our scope to Python---specifically the most popular packages as listed in \Cref{table:downloads} (a). 
We also observe that Bambi is a high-level wrapper built on PyMC, and TFP has superseded Edward, which is no longer actively developed \citep{bambi_docs_2025, tran2016edward}.

\begin{table}[t]
\centering
\footnotesize
\caption{Download counts (last 30 days, as of October~2025) for software packages that include variational inference functionality in  (\subref{table:downloads-python}) Python and (\subref{table:downloads-r}) R. }
\label{table:downloads}

\begin{subtable}[t]{0.48\linewidth}
\centering
\caption{Python packages}
\label{table:downloads-python}
\begin{tabular}{p{3.5cm}r}
\toprule
\textbf{Package} & \textbf{Last 30 days} \\
\midrule
CmdStanPy (\texttt{cmdstanpy}) & 9{,}444{,}333 \\
TensorFlow Probability (\texttt{tensorflow-probability}) & 1{,}283{,}307 \\
PyMC (\texttt{pymc}) & 927{,}688 \\
PyStan (\texttt{pystan}) & 599{,}396 \\
Pyro (\texttt{pyro-ppl}) & 560{,}959 \\
NumPyro (\texttt{numpyro}) & 518{,}507 \\
Bambi (\texttt{bambi}) & 52{,}496 \\
Edward2 (\texttt{edward2}) & 3{,}444 \\
Edward (\texttt{edward}) & 1{,}121 \\
\bottomrule
\end{tabular}
\end{subtable}
\hfill
\begin{subtable}[t]{0.48\linewidth}
\centering
\caption{R packages}
\label{table:downloads-r}
\begin{tabular}{lr}
\toprule
\textbf{Package} & \textbf{Last 30 days} \\
\midrule
\texttt{rstan} & 74{,}196 \\
\texttt{rstanarm} & 34{,}442 \\
\texttt{varbvs} & 1{,}002 \\
\texttt{tfprobability} & 988 \\
\texttt{greta} & 271 \\
\bottomrule
\end{tabular}
\end{subtable}

\end{table}

In particular, we examined the versions listed in \Cref{tab:versions-defaults} because they were the latest available in October 2025.

\begin{table}[t]\footnotesize\centering
\caption{Version of each software package studied in this manuscript.}
\label{tab:versions-defaults}
\begin{tabular}{lccc}
\toprule
Package & Version  \\
\midrule
PyMC & 5.25.1 \\ 
NumPyro & 0.18.0 \\
TFP & 0.25.0 \\
\bottomrule
\end{tabular}
\end{table}

\paragraph{Scope and Motivation}
In practice, PPLs with VI implementations rely on default choices---variational families, transformations for constrained parameters, initialization schemes, and optimizers---that we expect users rarely modify.\footnote{
This behavior mirrors the ``opt-out'' phenomenon: the tendency for most people to accept default options rather than take action to change them \citep{Davidai2012, Mehta2016}.
} Our goal is to characterize how these defaults behave in PyMC, NumPyro, and TFP, and to disentangle intrinsic limitations of VI from numerical or design effects specific to each software framework. 

\paragraph{Evaluating Simple Models}
While our broader interest lies in applying VI to complex hierarchical and deep models, we begin with univariate conjugate distributions as a minimal testbed. These settings drastically simplify the VI optimization problem, allowing us to isolate software-level numerical behavior.

Although our testbeds are simpler than the models we expect to see in practice, they can still represent realistic geometries that are close to those that arise in practice. For instance, the Bernstein--von Mises theorem shows that, under regularity conditions, posterior distributions become asymptotically Gaussian (as the number of data points increases) across many common choices of prior or likelihood (see \citealp[Ch.~10.2]{Vaart_1998}; \citealp{LeCam1953}). As a result, Gaussian posteriors can capture key behaviors of posteriors that arise in real-world applications.

We further focus on one-dimensional conjugate models; if an approximation fails in one dimension, we may reasonably expect the failure to occur or worsen in higher dimensions or more complex setups. 

\paragraph{One-Dimensional Gaussian--Gaussian (Unknown Mean, Known Variance)}
\label{sec:gaussian_gaussian}
Throughout this paper, we will use a one-dimensional conjugate Gaussian with known variance as our canonical test case. Given a known variance $\sigma^2$ and observations $Y$ drawn i.i.d.\ from a Gaussian likelihood with a Gaussian prior on the mean $\mu$, the posterior is also Gaussian with parameters $\mu_p, \sigma_p^2$, i.e., $p(\mu \mid Y) \sim \mathcal{N}(\mu_p, \sigma_p^2)$. Because the true posterior lies exactly within the Gaussian variational family, the optimal VI solution matches the true posterior. Moreover, a special case with $n=0$ simply recovers the prior. By having a known variational approximation, the one-dimensional Gaussian--Gaussian is a useful test for isolating software-induced VI errors. 
\FloatBarrier

\section{Results}\label{sec:results}

We evaluated PyMC, NumPyro, and TFP on the benchmark outlined at the end of \Cref{sec:rel_work}.
Under default settings in PyMC, we observe a  persistent systematic bias 
for millions of iterations. In all three software packages, we observe minor oscillations in variational parameter trajectories (i.e., the paths traced by the variational posterior parameters across iterations of the VI algorithm). Minor non-default adjustments---such as either increasing the number of Monte Carlo (MC) gradient samples or switching to Adam---resolve the systematic bias, and using both adjustments together reduces the magnitude of the oscillations in all cases.

We also observe differences across the software packages in how they handle initialization and transformations. Because typical VI objectives are non-convex, initialization can substantially affect the resulting posterior approximation. The software packages differ in how they expose control over initialization. The software packages also differ in how they apply and expose transformations. In the case of TFP, transformations are not automatically enacted, and VI may fail silently if a transformation is required: optimization may proceed, but the \ELBOspace value becomes \texttt{NaN} and the returned approximation is nonsensical. 

\subsection{Optimization Quality}

Here, we first describe undesirable behavior under the default single \mcnodenospace\ and default Adagrad\_Window optimizer in PyMC.  In PyMC's default settings, we observe a systematic bias of $\approx 5\%$ in the approximated variational posterior standard deviation for millions of iterations. We then discuss how to mitigate this behavior by changing software defaults, noting that while NumPyro and TFP's default behaviors are not problematic for our one-dimensional Gaussian-Gaussian benchmark, both software packages exhibit minor performance improvements when some defaults are changed as well.

\subsubsection{Software Implementation Setup}

Before describing the optimization behavior, we establish the default settings in PyMC, NumPyro, and TFP. We discuss which default settings are truly default, i.e., they will run without any user input, and which settings require user input to run without an error.

\paragraph{Number of Monte Carlo Gradient Samples (Default = 1)}\label{sec:mc1}
We establish that the default setting in PyMC, NumPyro, and TFP is to use a single \mcnodenospace. The package source code confirms these defaults directly. NumPyro sets \texttt{num\_particles=1} in its ELBO implementation \citep{NumPyro_elbo_2019}, TFP sets \texttt{sample\_size=1} in \texttt{fit\_surrogate\_posterior} 
\citep{tfp_vi_fit_surrogate_posterior_2024}, and PyMC uses \texttt{obj\_n\_mc=1} \citep{PyMC_obj_n_mc_advi_2018}. In all three packages, using a single MC sample is specified without requiring any user input. 

The single MC sample software package defaults may be informed by the claim in \citet{kucukelbir2016automatic} that using a single \mcnode per iteration in the numerical integration approximating the gradient of the \ELBOspace is sufficient in practice for desired performance. We will later see that, in practice, a single \mcnode does not always guarantee desirable behavior. 

\paragraph{Optimizer Choice: Adam vs. Adagrad\_Window}
PyMC does not require the user to specify an optimizer, as it automatically uses Adagrad\_Window. Both NumPyro and TFP require the user to specify an optimizer, and neither will run without a provided optimizer. Since many developer-written and officially released VI vignettes for NumPyro and TFP reference the Adam optimizer \citep{numpyro_autoDAIS, pyro_svi_part_i,  tfp_release_notebook_0_12_1, tfp_vi_fit_surrogate_posterior_2024, tfp_joint_distribution_2025, tfp_linear_mixed_effects_vi_2025}, we use Adam as the default for NumPyro and TFP. We summarize each package's default optimizer and whether the optimizer is run by default or must be chosen by the user in \Cref{tab:optimizer_choice}.

The default parameters of each optimizer used in this paper are displayed in \Cref{tab:optimizer_defaults}. We include PyMC's Adam optimizer alongside the default optimizers to demonstrate that small changes to the default PyMC behavior return results matching those of the other software packages' defaults. NumPyro's implementation of the Adam optimizer is a wrapper on Jax's Adam implementation \citep{numpyro_svi_latest_2025}. While PyMC, NumPyro, and TFP implement the $\beta_1$ and $\beta_2$ parameters identically as exponential decay rates for the first- and second-moment estimates in Adam, respectively, they differ in their incorporation of numerical stability constants (e.g., between NumPyro and TFP) \citep{keras_adam_optimizer_v3_3_3, jax_example_libraries_optimizers_2026}. 

\begin{table}[ht]
\centering
\begin{tabular}{lll}
\hline
\textbf{Package} & \textbf{Default optimizer} & \textbf{User-input vs.\ default} \\
\hline
PyMC 
& Adagrad\_Window 
& Default \\

NumPyro 
& Adam 
& User-input\\

TFP 
& Adam 
& User-input \\
\hline
\end{tabular}
\caption{Optimizers we use as default for each software package. Each is specified as either running by default without user intervention or requiring user specification of the optimizer. PyMC uses Adagrad\_Window as its true default optimizer without user intervention. In contrast, NumPyro and TFP examples and vignettes typically specify the use of the Adam optimizer \citep{numpyro_autoDAIS, pyro_svi_part_i,  tfp_release_notebook_0_12_1, tfp_vi_fit_surrogate_posterior_2024, tfp_joint_distribution_2025, tfp_linear_mixed_effects_vi_2025}}
\label{tab:optimizer_choice}
\end{table}

\begin{table}[ht]
\centering
\begin{tabular}{llp{3.4cm}p{4.3cm}}
\hline
\textbf{Package and Optimizer} & \textbf{step size} & \textbf{Other defaults} & \textbf{Notes} \\
\hline
PyMC Adagrad\_Window 
& $10^{-3}$ 
& $\epsilon_{\textrm{PyMC}} = 0.1$, $n_{\text{win}} = 10$ 
& True default optimizer used by PyMC \\

PyMC Adam 
& $10^{-3}$ 
& $\epsilon_{\textrm{PyMC}} = 10^{-8}$, $\beta_1=0.9$, $\beta_2 = 0.999$  
& -- \\

NumPyro Adam 
& -- 
& $\epsilon_{\textrm{NumPyro}} = 10^{-8}$, $\beta_1 = 0.9$, $\beta_2 = 0.999$  
& Wraps JAX Adam; user must provide step size\\

TFP Adam 
& $10^{-3}$ 
& $\epsilon_{\textrm{TFP}} = 10^{-7}$ , $\beta_1 = 0.9$, $\beta_2 = 0.999$, 
& -- \\
\hline
\end{tabular}
\caption{Default hyperparameters for optimizers we used in our experiments. All specified default values in the table are used by default without user input. NumPyro's Adam optimizer requires a user-specified step size and will fail with an error if none is provided. Following the NumPyro VI vignettes \citep{numpyro_docs_stable_2025, numpyro_svi_latest_2025}, we use $0.0005$ as the default step size for NumPyro Adam. }
\label{tab:optimizer_defaults}
\end{table}

\paragraph{Termination criteria}
In each software package, the default behavior is to terminate the algorithm at the specified number of iterations and return the current variational approximation as the reported variational posterior. If the number of iterations is left unspecified by the user, PyMC will terminate after $10{,}000$ iterations and both NumPyro and TFP will return an error stating that the user needs to choose a number of iterations for which to run the VI algorithm. 

While neither NumPyro or TFP will run without a provided number of iterations by the user, both software packages have vignettes that suggest different numbers of iterations for the VI algorithm. TFP's documentation and vignettes reference using $100$ iterations \citep{tfp_vi_fit_surrogate_posterior_2024}, $200$ iterations \citep{tfp_probabilistic_pca_2024}, $300$ iterations \citep{tfp_joint_distribution_2025}, $500$ iterations \citep{tfp_release_notebook_0_12_1}, $1000$ iterations \citep{tfp_joint_distribution_2025}, $3000$ iterations \citep{tfp_linear_mixed_effects_vi_2025}, and $10{,}000$ iterations \citep{tfp_variational_inference_joint_dist_2024}. NumPyro's documentation and vignettes reference using $2{,}000$ iterations \citep{numpyro_svi_latest_2025} and $5{,}000$ iterations \citep{pyro_svi_part_i}.

\subsubsection{Default Implementation Behavior}

\begin{figure}[p]
    \centering
    \includegraphics[width=0.9\linewidth]{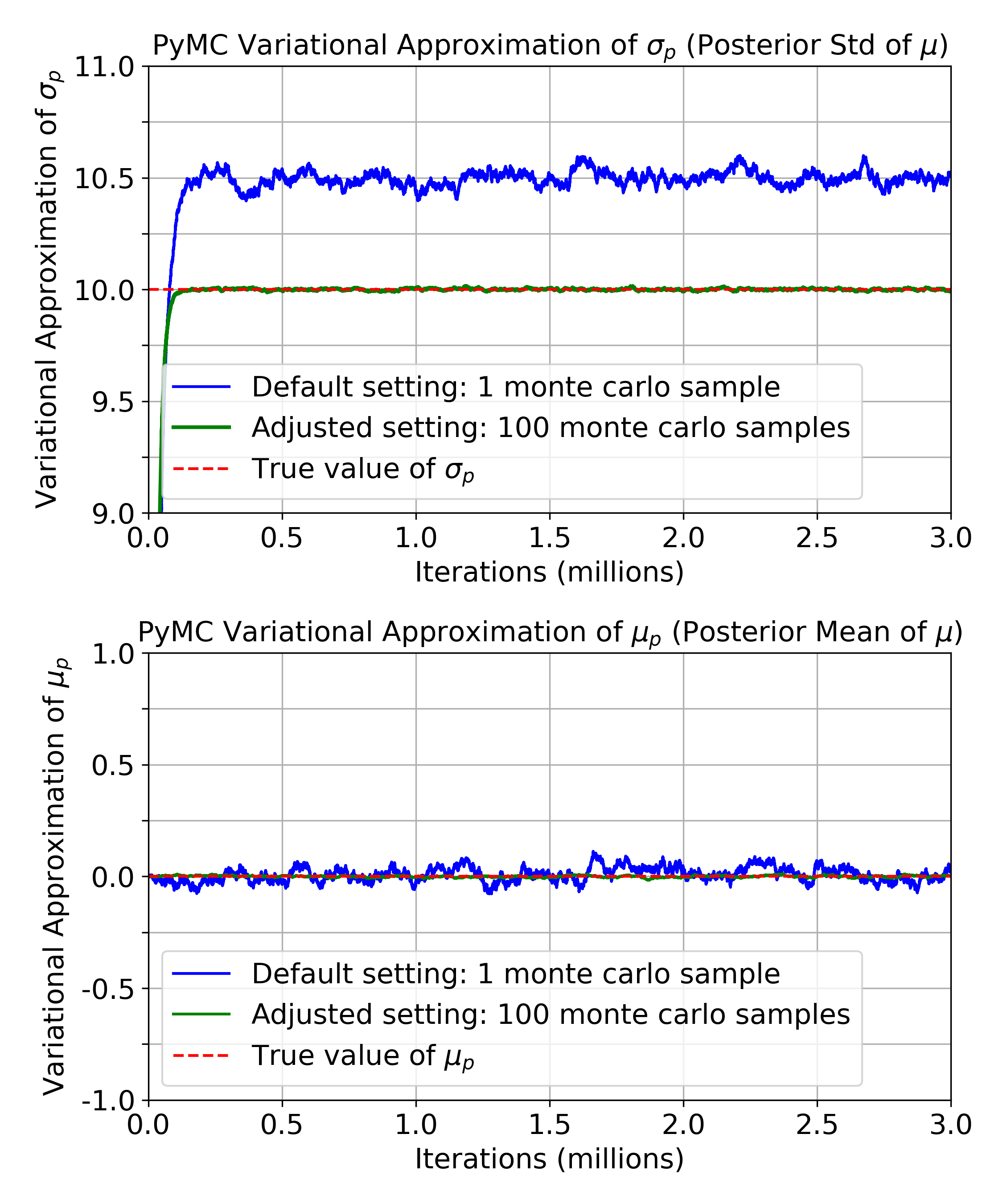}
    \caption[Default vs tuned \mcgradnodes in PyMC (one-dimensional Gaussian, known variance) with Adagrad\_ Window optimizer]{ Default vs. tuned \mcgradnodes in PyMC (one-dimensional Gaussian, known variance) run with PyMC's default Adagrad\_Window optimizer. The top and bottom figures plot the variational trajectories under default (1)  and adjusted (100) \mcnode values for the posterior parameters $\sigma_p$ and $\mu_p$, respectively, where trajectories of the same color between plots are derived from the same VI run. 

    \textbf{Upper plot:} Using a single \mcnode overshoots and settles at a biased value; increasing the number of \mcnodes to 100 causes the variational approximation of $\sigma_p$ to settle at the analytic posterior standard deviation (dashed line).
    \textbf{Lower plot:} Using a single \mcnode (blue) results in minor trajectory oscillations; increasing the number of \mcnodes to 100 (green) reduces the magnitude of oscillatory behavior.
    }
    \label{fig:PyMC_prelim}
\end{figure}

The top plot of \Cref{fig:PyMC_prelim} visualizes the trajectory of the approximate posterior standard deviation $\sigma_p$ on the one-dimensional Gaussian--Gaussian benchmark run in PyMC. We use default settings: one \mcnode  and Adagrad\_Window. Model details are in \Cref{sec:gaussian_gaussian}.   The trajectory of $\sigma_p$ with 1 MC sample (blue line) overshoots and settles at a posterior standard deviation about $5\%$ above the analytic target (red dashed line). Considering the approximate posterior mean $\mu_p$, instead, yields the bottom plot of \Cref{fig:PyMC_prelim} displays the PyMC trajectory of $\mu_p$ in the same run of the algorithm as in the top plot. We see small-magnitude oscillations and are not concerned with this default behavior. 

In conjugate beta, inverse-gamma, and gamma models, outlined in \Cref{sec:benchmarks_conjugate}, we expect to observe at least as much systematic bias in the approximated variational posterior standard deviation when using PyMC with default settings. Our preliminary further experiments suggest that the bias does persist.

While it is well-noted in the literature that stochastic gradient methods may struggle to clearly converge when used with ADVI \citep{Giordano2024BlackBox} and Adam is a more robust optimizer of the ELBO \citep{campbell_placeholder}, to our knowledge, this particular bias issue when using PyMC's defaults has not been previously documented in such simple cases.

Next, we consider NumPyro and TFP, with their default settings: one \mcnode and Adam. In NumPyro and TFP, when we run both with the optimizers in \Cref{tab:optimizer_choice} and optimizer settings from \Cref{tab:optimizer_defaults}, trajectories for both $\mu_p$ and $\sigma_p$ demonstrate minor oscillations that we do not consider to be problematic and no systemic bias. The oscillation behaviors of all three software packages are consistent with previous observations for low–sample stochastic VI \citep{Giordano2024BlackBox}. 

We expect to see similar behaviors for 
conjugate beta, inverse-gamma, and gamma models. Again, our preliminary further experiments suggest the bias persists.

The various plots and the iteration-by-iteration tracked values illustrated in this paper are not provided by default in any of the software implementations.

\begin{figure}[p]
    \centering
    \includegraphics[width=0.9\linewidth]{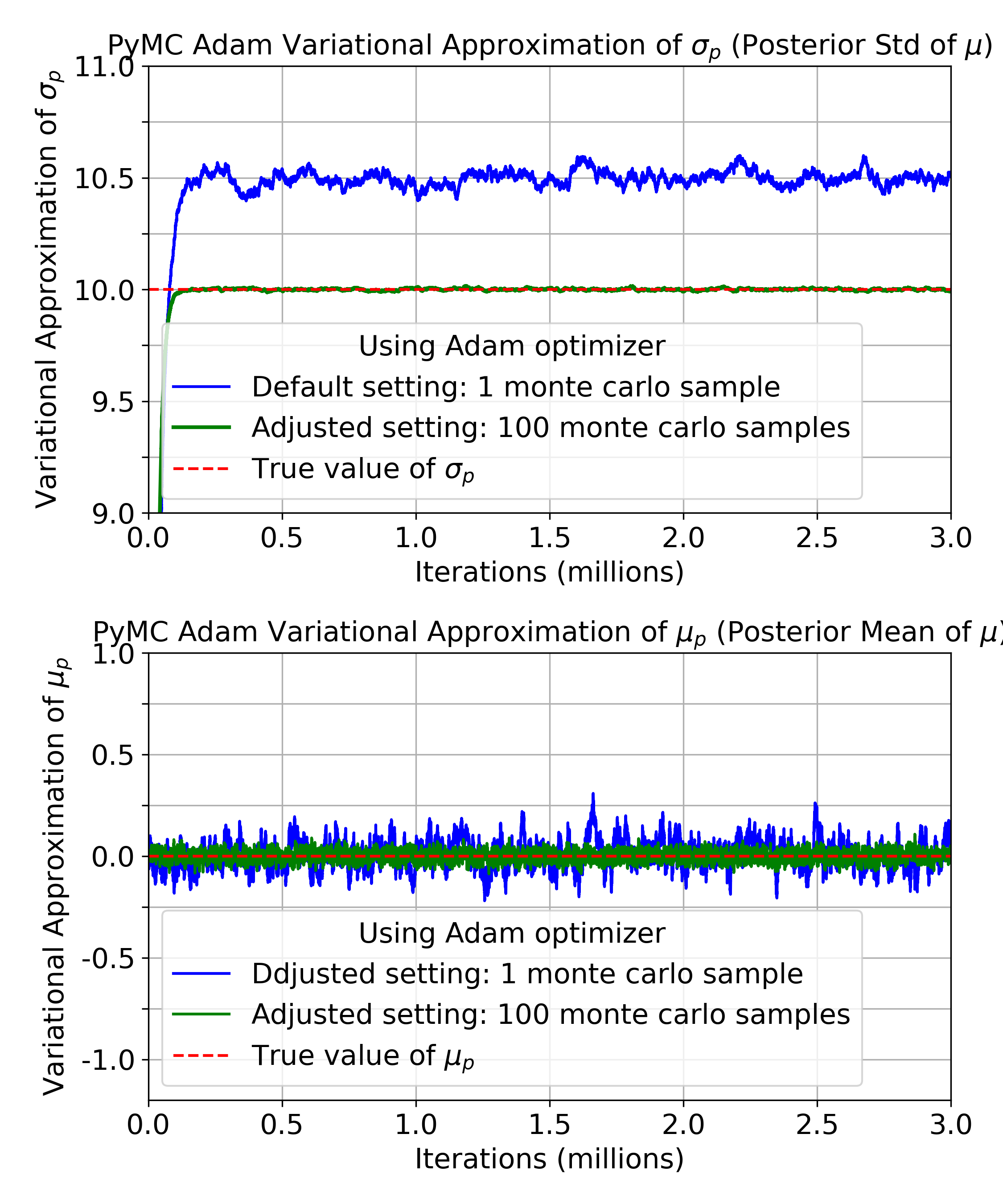}
    \caption[Default vs tuned \mcgradnodes in PyMC (one-dimensional Gaussian, known variance) with Adam optimizer]{ Default vs. tuned \mcgradnodes in PyMC (one-dimensional Gaussian, known variance) run with Adam optimizer and default Adam settings. The upper and lower figures plot the variational trajectories under default (1)  and adjusted (100) \mcnode values for the posterior parameters $\sigma_p$ and $\mu_p$, respectively, where trajectories of the same color between plots are derived from the same VI run. 

    \textbf{Upper plot:} Using a single \mcnode displays minor oscillations around analytic posterior standard deviation (dashed line); increasing the number of \mcnodes to 100 further decreases oscillations.
    \textbf{Lower plot:} Using a single \mcnode (blue) results in minor trajectory oscillations; increasing the number of \mcnodes to 100 (green) reduces the magnitude of oscillatory behavior.
    }
    \label{fig:PyMC_adam}
\end{figure}

\begin{figure}
    \centering
    \includegraphics[width=0.9\linewidth]{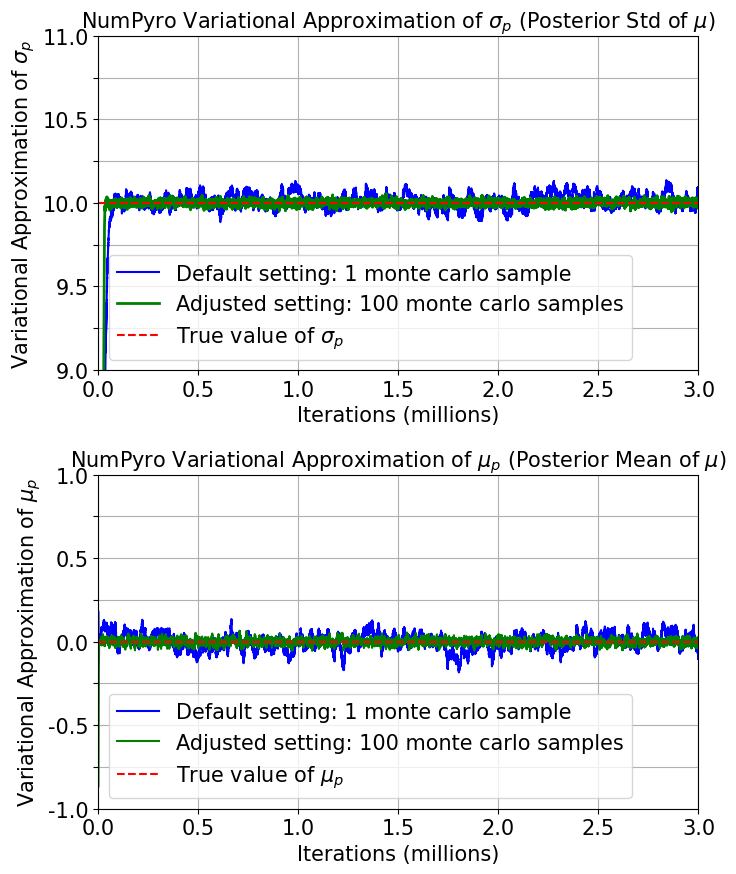}
    \caption[Default vs tuned \mcgradnodes in NumPyro (one-dimensional Gaussian, known variance)]{Default vs. tuned \mcgradnodes in NumPyro (one-dimensional Gaussian, known variance). The top and bottom figures plot the variational trajectories under default (1)  and adjusted (100) \mcnode values for the posterior parameters $\sigma_p$ and $\mu_p$, respectively, where trajectories of the same color between plots are derived from the same VI run. 

    \textbf{Upper plot:} Using a single \mcnode oscillates with a small magnitude; increasing the number of \mcnodes to 100 decreases the oscillation around the analytic posterior standard deviation (dashed line).
    \textbf{Lower plot:} Using a single \mcnode (blue) results in small trajectory oscillations; increasing the number of \mcnodes to 100 (green) reduces the magnitude of oscillatory behavior.}
    \label{fig:NumPyro_prelim}
\end{figure}

\begin{figure}
    \centering
    \includegraphics[width=.92\linewidth]{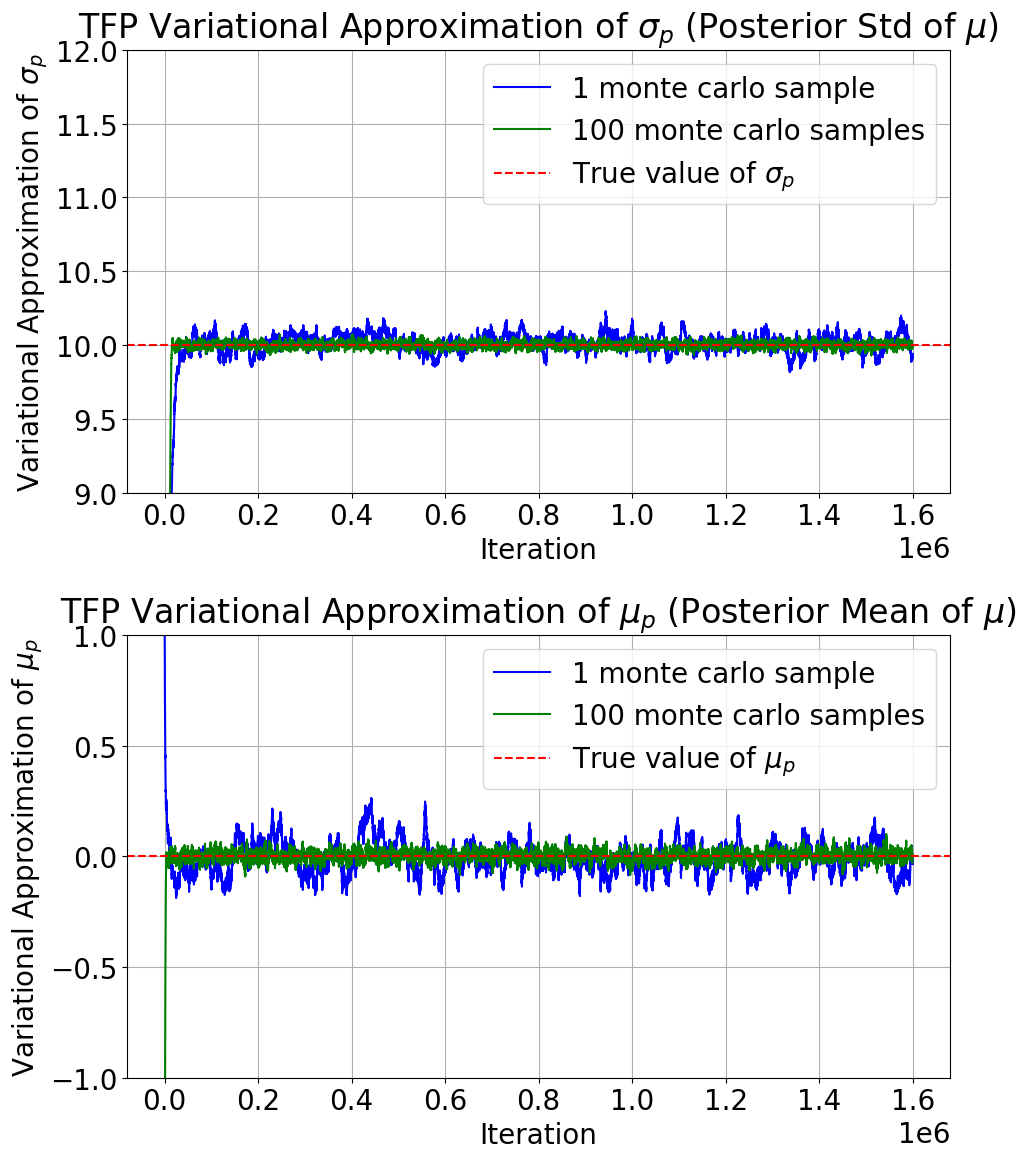}
    \caption[Default vs tuned \mcgradnodes in TFP (one-dimensional Gaussian, known variance)]{Default vs. tuned \mcgradnodes in TFP (one-dimensional Gaussian, known variance). The top and bottom figures plot the variational trajectories under default (1)  and adjusted (100) \mcnode values for the posterior parameters $\sigma_p$ and $\mu_p$, respectively, where trajectories of the same color between plots are derived from the same VI run. 

    \textbf{Upper plot:} Using a single \mcnode oscillates with a small magnitude; increasing the number of \mcnodes to 100 decreases the oscillation around the analytic posterior standard deviation (dashed line).
    \textbf{Lower plot:} Using a single \mcnode (blue) results in oscillatory trajectories that are larger than those produced after increasing the number of \mcnodes to 100 (green).}
    \label{fig:TFP_prelim}
\end{figure}

\subsubsection{Behavior Mitigation and Improvement}

We find that PyMC's systemic bias can be mitigated by changing the default values, and all of the non-problematic oscillations can also be improved by changing the default values. For PyMC, we see in the top of \Cref{fig:PyMC_prelim} that the trajectory of $\sigma_p$ under increased \mcnode count ($100$ MC samples, green line) settles at the best variational approximation. If we instead use Adam as the optimizer, as demonstrated in \Cref{fig:PyMC_adam}, we observe that the resulting trajectory of $\sigma_p$ is unproblematic with 1 MC sample (blue line) and may be considered to be improved with 100 MC samples (green line).

Also for PyMC, we see at the bottom of \Cref{fig:PyMC_prelim} that the trajectory of $\mu_p$ under increased \mcnode count (green line) has a smaller magnitude of oscillations around the analytic ground truth posterior value. When we change the optimizer to Adam, the resulting trajectory of $\mu_p$ remains acceptable and behaves much the same as with the Adagrad\_Window optimizer.

We observed unconcerning behavior for both parameters in NumPyro and TFP; both $\mu_p$ and $\sigma_p$ exhibit smaller oscillations around the analytic ground-truth posterior value when run with increased \mcnode count compared to the default of a single MC sample. We see the default and slightly improved behavior in NumPyro in \Cref{fig:NumPyro_prelim} and observe similar behavior in TFP in  \Cref{fig:TFP_prelim}. With the exception of \Cref{fig:PyMC_adam}, where we run PyMC with the Adam optimizer, we run each figure in the respective software package with its default optimizer and default optimizer parameters, for a manual input of $1.6$ million iterations, using the same prior and likelihood parameters as in the PyMC case above.

While the 1 MC sample oscillations in NumPyro, TFP, and the $\mu_p$ parameter of PyMC are not concerning in the one-dimensional Gaussian-Gaussian benchmark, we do note that increasing the number of MC samples further decreases the magnitude of oscillations, indicating that all three software packages stand to be improved by the slight change in defaults of increasing the number of default MC samples.

\subsection{Initialization}\label{sec:init}

Results from VI software can be sensitive to initialization.  NumPyro and TFP both offer random restarts by default without a set seed, so multiple default runs yield different initializations. PyMC, however, has a fixed default initialization.
All three software packages provide non-default options that enable users to explore different initializations.

\paragraph{Initialization Sensitivity}
We begin by describing the importance of initialization.
In variational inference, the objective may be non-convex \citep{blei2017variational}, and the reverse-KL divergence that is commonly used in VI is empirically mode-seeking, often capturing only a single mode when the posterior is multimodal \citep[sec.~2.1]{jerfel2021variational}. 
 
 Because the \ELBOspace may exhibit multiple local optima,  the initialization of the variational parameters can materially affect the attained optima. As a result, we may reasonably expect multiple distinct initializations to aid in better exploring the totality of posterior modes.

\paragraph{TFP initialization behavior}
By default, TFP initializes the unconstrained mean \(\underlyingmu\) by drawing from a \(\mathcal{N}(0,1)\) distribution and initializes the unconstrained standard deviation \(\underlyingsigma\) by setting it to $0.01$.  TFP does not set a random seed by default, so repeated runs of the same VI procedure should yield different initializations, encouraging exploration of the posterior support.

Beyond the default functionality, TFP also allows users to explicitly initialize both the unconstrained mean \(\underlyingmu\) and standard deviation \(\underlyingsigma\) of the underlying normal distribution; they may be set arbitrarily by the user.

\paragraph{NumPyro default initialization behavior}
By default, NumPyro uses a draw from a \(\mathrm{Uniform}(-2,2)\) distribution to set the initial unconstrained mean \underlyingmu\ and sets \(\underlyingsigma = 0.1\).   As with TFP, NumPyro does not fix a random seed by default, so repeated runs should yield different initializations.

Beyond the default, NumPyro offers several predefined initialization functions for the unconstrained mean \underlyingmu, such as initializing it to the median of a collection of prior samples in the unconstrained space or specifying a fixed value. The user can also numerically set the initial value of \underlyingsigma. Concretely, users may provide an \texttt{init\_loc\_fn} to determine the initial value of \underlyingmu\ and an \texttt{init\_scale} to control the initial value of \underlyingsigma.

\paragraph{PyMC initialization behavior}
By default, PyMC initializes the unconstrained mean \underlyingmu\ to a fixed value given the specified model. For any choice of prior distribution, PyMC provides a predefined starting value, denoted \texttt{support\_point}. PyMC selects a transformation \(T\) based on the prior and likelihood's implied support for the posterior. It applies the inverse transformation \(T^{-1}\) to the starting value, assigning the result as the initial value of the unconstrained mean \(\underlyingmu\).
PyMC sets the unconstrained standard deviation \(\underlyingsigma\) to a fixed default value of approximately \(0.693\), corresponding to the softplus inverse of 1 \citep{pymc_repo_2025}. 

Beyond the default initialization behavior, PyMC allows users to explicitly specify the \texttt{support\_point} value via a start dictionary, as discussed in \citet{PyMC_how_to_initialize_advi_2023}. PyMC then proceeds with the same transformation procedure as in the default case to choose the \underlyingmu\ value. PyMC does not offer initialization options to change the unconstrained standard deviation \underlyingsigma.

\begin{figure}
    \centering
    \includegraphics[width=\linewidth]{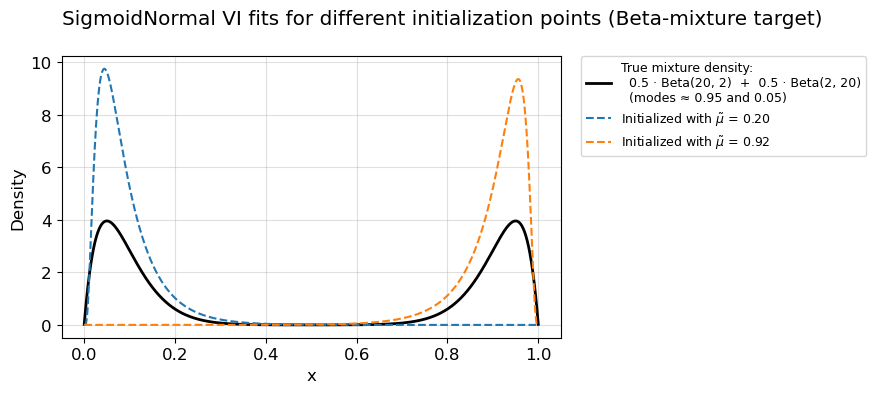}
    \caption[SigmoidNormal VI approximations to a mixture of betas with two narrow modes]{ The true target density (black solid line) is a balanced mixture of $\textrm{beta}(20, 2)$ and $\textrm{beta}(2, 20)$, producing two narrow modes. Shown are variational approximations (dashed lines) obtained by fitting a Gaussian in the unconstrained space and mapping it to $(0, 1)$ via a sigmoid transformation using the NumPyro package with a default value of $\tilde{\sigma}=0.1$. Each color corresponds to a different initialization of \underlyingmu.   }\label{fig:beta_mix_init}
\end{figure}

\paragraph{Multimodality Concerns}
In multimodal models, initialization can affect which mode VI captures. Our example in \Cref{fig:beta_mix_init} illustrates that VI can become trapped in a local optimum under specific initializations. In particular, we set up a density (black solid line) that corresponds to a bimodal distribution not only in the constrained space but also in the unconstrained space.\footnote{
We note that we need a mixture of two betas rather than a single beta distribution to achieve bimodality in the unconstrained space. While a single beta can be bimodal in the \emph{constrained} space with an appropriate choice of parameters, its transformed distribution to \emph{unconstrained} space is strictly unimodal. See \Cref{sec:beta_bernoulli} for model details.}
Across different initializations of the variational parameters, reverse-KL variational inference converges to distinct local optima, each of which concentrates mass near a single mode of the target. Each dashed curve corresponds to a different initialization, illustrating the strong sensitivity of the optimization outcome to initialization in this setting. 

The mode-seeking behavior need not always be a problem, though.
In many clustering settings, label non-identifiability means that our desired outcome is to recover the posterior corresponding to a single labeling of the clusters (not the posterior averaged over all labelings). The importance of multiple initializations depends on the application.

\subsection{Control and Defaults of Transformations}

Each software package offers different control and types of transformations. NumPyro and PyMC both automatically apply transformations derived from defined prior support constraints (noted in \Cref{fig:NumPyro_transformation_table} and \Cref{fig:PyMC_transformation_table}). The transforms from NumPyro and PyMC cannot be overridden without changing the prior distribution or the support constraints, respectively. By default, TFP does not automatically apply any transforms; users must explicitly add bijectors. 
Further, we observe that when a transformation is needed and not specified by the user, TFP can fail silently.

\paragraph{Software Implementation Details} In TFP, while users can choose to add transformations to their variational family, the default is not to have a transformation. If there are any support constraints, TFP will, by default, run VI without using a transformation and not raise any errors, but it will have a support mismatch. NumPyro and PyMC both automatically choose transformations from the defined prior support constraints 
\citep{PyMC_transforms_api_2025, NumPyro_github_2025}. We note that the user cannot override the selected transformations in NumPyro or PyMC. See \Cref{sec:softplus} for details on each of the transformations used by the software packages.

\begin{figure}[ht!]
    \centering
    \includegraphics[width=1\linewidth]{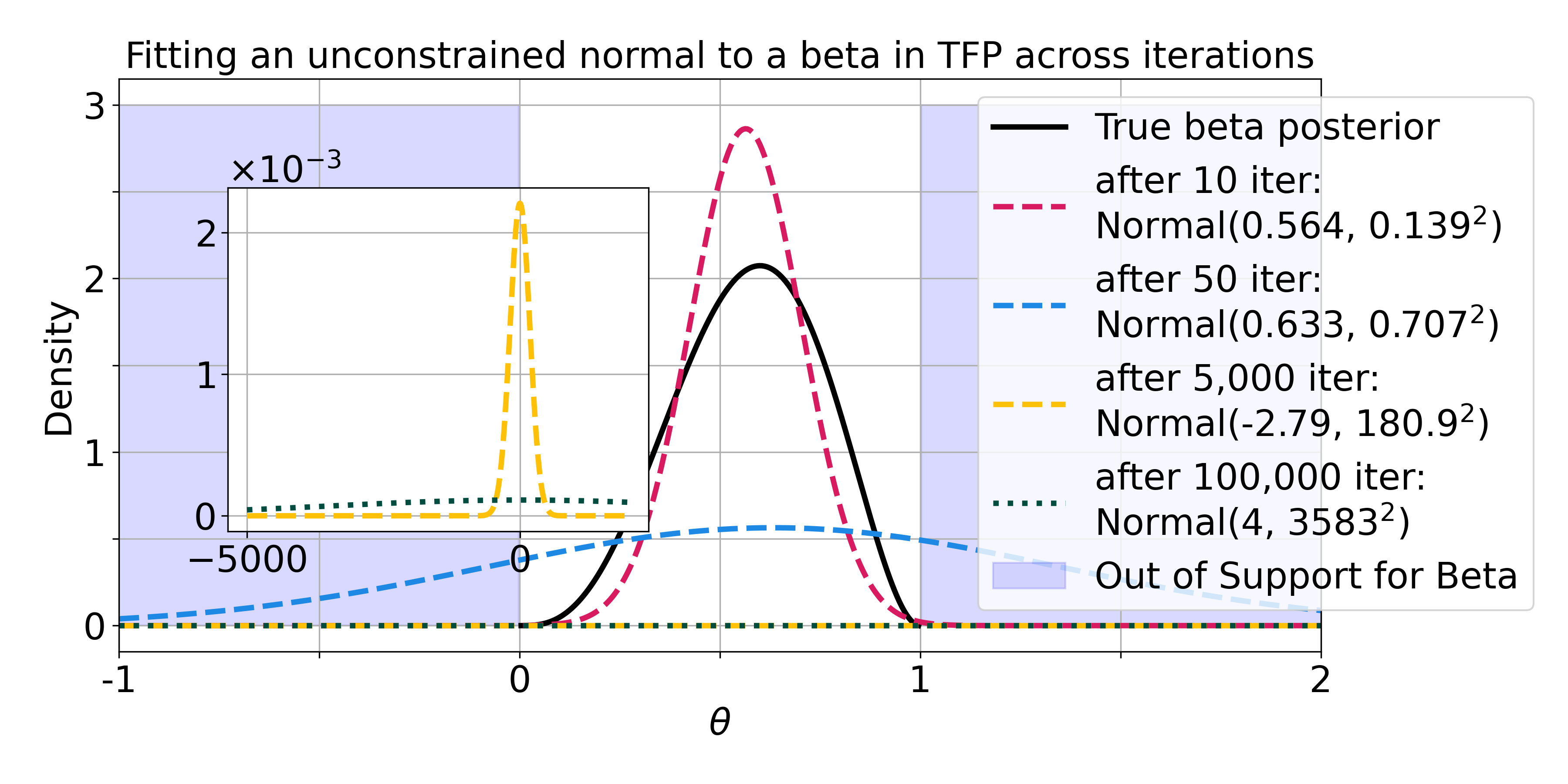}
    \caption[Support mismatch without a transform]{Support mismatch without a transform. The untransformed fit drifts out of support. Adding the sigmoid bijector yields a valid fit. The inset axes zoom out on the fitted VI distributions after 5,000 and 100,000 iterations to show behavior not visible on the smaller scale of the main plot.

    }

    \label{fig:non_transformed}
\end{figure}

\paragraph{Implementation Behavior}
In \Cref{fig:non_transformed}, we compare fitting an untransformed Gaussian variational family to fitting a transformed Gaussian variational family to a conjugate beta distribution (detailed in \Cref{sec:beta_bernoulli}), demonstrating how the ill-definition of the \ELBOspace creates a poor approximation of the conjugate beta without the proper transformation. We plot four normal approximations, run in TFP on the same seed, after 10, 50, 5{,}000, and 100{,}000 iterations, respectively. As TFP's VI algorithm runs for more iterations, the untransformed normal distribution drifts into the positive reals, producing an invalid MC approximation to the \ELBO\ and an unusable approximation. 
 Adding the appropriate bijector (e.g., a sigmoid) restores support validity and produces a meaningful fit.

\paragraph{Why Transformations Matter}
Without a support-matching transformation, an untransformed Gaussian variational family can assign probability mass outside the posterior’s domain (e.g., outside of the unit interval for a beta distribution).

\subsection{Diagnostics and Tracking using Transformations}

 To identify whether VI is encountering difficulties, diagnostics can help detect oscillations, premature plateaus, stabilization rates, and sample efficiency. However, monitoring only the unconstrained parameters can be misleading: large fluctuations in the unconstrained space may correspond to negligible changes after transformation, especially near constraint boundaries where nonlinear distortions compress scale. Because the constrained space carries the interpretable units of the original model, and we want to report posterior summaries there, we prefer it and believe it may be helpful to have diagnostics in the constrained space after transformation, where evaluation metrics, analytical references, and decision thresholds are defined.

While none of the three software packages provides an explicit report of the mean and variance of the variational parameters in the constrained space during VI fitting in their default VI interfaces explored in this paper, they all provide access to parameter tracking on the unconstrained scale.

\paragraph{Software Implementation Details}

In TFP, users can access \underlyingmu\ and \underlyingsigma, the mean and standard deviation of the underlying normal distribution of the variational approximation, by calling the \texttt{loc} and \texttt{scale} attributes of our variational approximation of the posterior. In NumPyro, users can access \underlyingmu\ and \underlyingsigma\ through the \texttt{get\_params()} method. In PyMC, users can access \underlyingmu\ and \underlyingsigma\ by calling \texttt{advi.approx.mean.eval} and \texttt{advi.approx.std.eval} methods, respectively. 

All three software packages provide access to constrained-space posterior means and standard deviations at the end of VI, but none provide a way to access them during optimization.

\paragraph{Diagnostics and tracking across spaces.}
Because transformations $T$ are often nonlinear, small changes in constrained parameters (e.g., probabilities near the boundary of $(0,1)$) may correspond to extremely large changes on the unconstrained scale, and vice versa. Monitoring and comparing within the constrained space avoids scale distortions and mismatches and directly reflects the posterior quantities practitioners report and evaluate.

In all three packages, since we do not have existing access to the current constrained parameters during optimization, the user must instead implement their own iteration-by-iteration technique; we used a method---known as measure transport---where we sample from the unconstrained Gaussian, transform via the known transformation $T$, and compute constrained summaries from the transformed samples \citep{pushforward}.
We note that using the measure transport method introduces additional Monte Carlo variability that diminishes with increasing numbers of MC draws used to approximate constrained parameters.

\FloatBarrier

\section{Practitioner User Guide}\label{sec:prac_guide}

This section distills our evaluation into risk-aware guidance for practitioners who wish to use VI in PyMC, NumPyro, and TFP on problems similar to our benchmarks. We summarize settings that reduced specific failure modes in the examples we examined. We do not expect these settings will eliminate all failure modes, but we believe that these instructions will be helpful in general. 

\subsection{Monte Carlo Gradient Estimation}

\paragraph{Recommendation} Increase Monte Carlo gradient samples to at least 100. The variables \texttt{num\_particles} (NumPyro), \texttt{sample\_size} (TFP),  \texttt{obj\_n\_mc} (PyMC) control the number of \mcnodes used in each respective software implementation. In each software package, set the respective variable to at least $100$.

\paragraph{Rationale} More \mcnodes reduce stochastic gradient variance, which reduces bias/oscillation in our tests. In our benchmark,  the default $1$ MC sample yielded $\approx5\%$ variance bias or small persistent oscillations; increasing to $100$ \mcnodes removed the bias and reduced the oscillations. We expect additional MC samples to be helpful in other models because using more MC samples will reduce the variance in the gradient estimate.

\paragraph{Possible Tradeoff} More MC samples will incur additional runtime cost.

\subsection{Optimizer and Learning Rate Choices}
\paragraph{Recommendation} Switch to using Adam if using PyMC. PyMC is the only software package that we tested that does not default to Adam. Within the \texttt{advi.fit} call in PyMC, use \texttt{obj\_optimizer=pm.adam(learning\_rate)} to change the optimizer to Adam.

\paragraph{Rationale} In our PyMC tests, switching from Adagrad\_Window to Adam on the same settings resolved the systematic bias error  (\Cref{sec:mc1}).  In non-convex objectives, Adagrad\_Window can prematurely shrink steps and stall progress \citep{wilson2017marginal}, while Adam is able to adapt to recent curvature and may be more reliable on VI objectives \citep{kingma2015adam, reddi2018convergence, campbell_placeholder}. 

\subsection{Initialization}
\paragraph{Recommendation} Use multiple random restarts or manual initialization when you can. If you would like to control initialization values of \underlyingmu\ and \underlyingsigma\ numerically, use NumPyro or TFP. PyMC does not allow initialization of the unconstrained standard deviation \underlyingsigma\ but users may influence the initialization of the unconstrained mean \underlyingmu. If you're using PyMC, you can set the starting value for the parameter of interest using the \texttt{start} parameter. If you're using NumPyro, you can set the initial parameters using \texttt{init\_loc\_fn} and \texttt{init\_scale}.  If you're using TFP, you can change the \texttt{initial\_parameters}. 

\paragraph{Rationale} For our one-dimensional unimodal benchmark, default initializations typically found the single mode; when they didn’t, the issue wasn’t initialization (\Cref{fig:non_transformed}). If you're working in multimodal settings where finding multiple modes is desirable, we recommend multiple random restarts.

\subsection{Transformations}
\paragraph{Recommendation} Confirm that, when working with a constrained support, a transform is being implemented to map from the unconstrained reals to the model's support and that reported variational parameters are in the constrained space.

\paragraph{Rationale} Omitting transformations can yield invalid approximations. Mismatched supports can produce out-of-support optima and unstable/ill-conditioned \ELBOspace estimates. In our beta example (\Cref{fig:non_transformed}), omitting the sigmoid bijector caused drift and an unstable fit.

\subsection{Summary}

Across all three frameworks, defaults sometimes yielded acceptable results in our one-dimensional tests. Our recommendations focus on users who use the package defaults. For such a user, after our analysis, we recommend using NumPyro and double-checking the support constraints. We prefer NumPyro because it has not demonstrated bias like PyMC or silent failure without transformations like TFP. NumPyro also requires explicit constraint specifications, which reduces tracking ambiguity and support mismatches in our tests. We like the level of control offered by NumPyro and the ease of debugging and diagnostics. For a default user, we recommend not using PyMC or TFP since they both may fail without changing default settings.
Following our recommendations in this subsection, the sensitivities we observe (bias, mode-sticking, and support mismatch) are reduced, and the optimization behavior becomes more predictable in our one-dimensional benchmarks.

\FloatBarrier

\section{Conclusion and Limitations}
\label{sec:conclusion}

\subsection{Summary of Findings}
Our findings reveal a clear pattern: accuracy under default settings is not guaranteed and depends on increasing the number of Monte Carlo gradient samples, using Adam, and verifying support-matching transforms.

\subsection{Implications for Developers}
Our findings also carry lessons for PPL developers:
\begin{itemize}
    \item Create vignettes with consistent parameter values to make developer recommendations clear.
    \item Implement default transformations.
    \item Change the default number of \mcnodes to $100$.
    \item Default to using the Adam optimizer.\footnote{The work of \citet{campbell_placeholder} also supports the suggestion of using the Adam optimizer.}
    \item Add constrained-space tracking methods.
\end{itemize}
Such improvements would make VI more trustworthy for new users and more reliable in production workflows.

\subsection{Limitations}

\subsubsection{Why We Need to Test More Models}
Our evaluation focuses on one-dimensional conjugate cases (especially the Gaussian--Gaussian conjugate model) to ensure a well-understood variational approximation is available and to ensure we can visualize the entire distribution. However, most applied models are non-conjugate and have more than one dimension.

Even one-dimensional models beyond our Gaussian benchmark with known variational approximations would allow us to explore the impact of transformations while having a known variational approximation to compare VI results to for verification of correctness.

A priori, we might expect conjugate models to be especially simplistic. Non-conjugate models (including widely used hierarchical models) may have distinctive posterior geometries, underscoring the risk of overgeneralizing results from conjugate benchmarks and highlighting the importance of expanding future evaluations to include such models.

Moreover, software packages sometimes have different implementations when analytic posteriors exist; in NumPyro, for example, there are two ELBO estimators: \texttt{Trace\_ELBO} estimator is used for non-conjugate models, while \texttt{TraceMeanField\_ELBO} takes advantage of analytic KL terms for conjugate cases. We observe that \texttt{TraceMeanField\_ELBO} performs substantially better when conjugacy holds. Since this paper's results are limited to conjugate cases, new issues may arise in non-conjugate cases.

\subsubsection{Scope and Generalizability}

Our scope is intentionally narrow, focusing on particular one-dimensional conjugate baselines in order to isolate the effects of VI implementation. 
As a next step, we will expand our evaluation to a wider selection of conjugate baselines. We will then extend to non-conjugate and hierarchical models where no analytic ground truth is available, high-dimensional and heavy-tailed models where optimization is more challenging, and runtime and scaling comparisons across frameworks to complement our focus on accuracy. We plan to extend our analysis to include the Stan framework to more comprehensively represent the most popular software packages offering VI. We're particularly interested in whether our recommendations still hold for more complex models. Extending the evaluation framework in these directions will provide a greater understanding of how VI defaults behave in practice and guide both users and developers toward more robust inference.

See our demo code at \url{https://github.com/madelynandersen/simpleVI/}.

\FloatBarrier

\cleardoublepage

\bibliography{refs}

@book{hoff2009firstch4,
  edition = {1st},
  title = {A {First} {Course} in {Bayesian} {Statistical} {Methods}},
  publisher = {Springer Publishing Company},
  author = {Hoff, Peter D.},
  month = jun,
  year = {2009},
  
}

@article{kucukelbir2016automatic,
    author  = {Alp Kucukelbir and Dustin Tran and Rajesh Ranganath and Andrew Gelman and David M. Blei},
  title   = {Automatic Differentiation Variational Inference},
  journal = {Journal of Machine Learning Research},
  year    = {2017},
  volume  = {18},
  number  = {14},
  pages   = {1--45},
}

@article{giordano2016fast,
  title={Fast robustness quantification with variational Bayes},
  author={Giordano, Ryan and Broderick, Tamara and Meager, Rachael and Huggins, Jonathan and Jordan, Michael},
  journal={Proceedings of the 2016 ICML Workshop on \#Data4Good: Machine Learning in Social Good Applications},
  year={2016},
}

@article{jordan1999introduction,
  title = {An {Introduction} to {Variational} {Methods} for {Graphical} {Models}},
  journal = {Machine Learning},
  volume = {37},
  number = {2},
  pages = {183--233},
  author = {Jordan, Michael I. and Ghahramani, Zoubin and Jaakkola, Tommi S. and Saul, Lawrence K.},
  year = {1999},
}

@article{giordano2018covariances,
 author  = {Ryan Giordano and Tamara Broderick and Michael I. Jordan},
  title   = {Covariances, Robustness, and Variational Bayes},
  journal = {Journal of Machine Learning Research},
  year    = {2018},
  volume  = {19},
  number  = {51},
  pages   = {1--49},
}

@article{strumbelj2024past,
  title = {Past, {Present} and {Future} of {Software} for {Bayesian} {Inference}},
  volume = {39},
  number = {1},
  journal = {Statistical Science},
  author = {Štrumbelj, Erik and Bouchard-Côté, Alexandre and Corander, Jukka and Gelman, Andrew and Rue, Håvard and Murray, Lawrence and Pesonen, Henri and Plummer, Martyn and Vehtari, Aki},
  month = feb,
  year = {2024},
  pages = {46--61},
}

@InProceedings{ranganath2014black,
  title = 	 {{Black Box Variational Inference}},
  author = 	 {Ranganath, Rajesh and Gerrish, Sean and Blei, David},
  booktitle = 	 {Proceedings of the Seventeenth International Conference on Artificial Intelligence and Statistics},
  pages = 	 {814--822},
  year = 	 {2014},
  editor = 	 {Kaski, Samuel and Corander, Jukka},
  volume = 	 {33},
  series = 	 {Proceedings of Machine Learning Research},
  address = 	 {Reykjavik, Iceland},
  month = 	 {22--25 Apr},
  publisher =    {PMLR},
}

@article{blei2017variational,
  title = {Variational {{Inference}}: {{A Review}} for {{Statisticians}}},
  shorttitle = {Variational {{Inference}}},
  author = {Blei, David M. and Kucukelbir, Alp and McAuliffe, Jon D.},
  year = {2017},
  month = apr,
  journal = {Journal of the American Statistical Association},
  volume = {112},
  number = {518},
  eprint = {1601.00670},
  primaryclass = {stat},
  pages = {859--877},
  archiveprefix = {arXiv},

}

@article{carpenter2017stan,
  author={Carpenter, Bob and Gelman, Andrew and Hoffman, Matthew D. and Lee, Daniel and Goodrich, Ben and Betancourt, Michael and Brubaker, Marcus and Guo, Jiqiang and Li, Peter and Riddell, Allen},
  title={Stan: A Probabilistic Programming Language},
  journal={Journal of Statistical Software},
  year={2017},
  volume={76},
  number={1},
  pages={1--32},
}

@article{salvatier2016probabilistic,
  author={Salvatier, John and Wiecki, Thomas V. and Fonnesbeck, Christopher},
  title={Probabilistic Programming in Python Using PyMC3},
  journal={PeerJ Computer Science},
  year={2016},
  volume={2}
}

@article{bingham2019pyro,
  author  = {Eli Bingham and Jonathan P. Chen and Martin Jankowiak and Fritz Obermeyer and Neeraj Pradhan and Theofanis Karaletsos and Rohit Singh and Paul Szerlip and Paul Horsfall and Noah D. Goodman},
  title   = {Pyro: Deep Universal Probabilistic Programming},
  journal = {Journal of Machine Learning Research},
  year    = {2019},
  volume  = {20},
  number  = {28},
  pages   = {1--6},
}

@article{wang2019frequentist,
  title={Frequentist Consistency of Variational Bayes},
  volume={114},
  number={527},
  journal={Journal of the American Statistical Association},
  author={Wang, Yixin and Blei, David M.},
  year={2019},
  pages={1147--1161}
}

@article{zhang2019advances,
  author={Zhang, Cheng and Butepage, Judith and Kjellstrom, Hedvig and Mandt, Stephan},
  title={Advances in Variational Inference},
  journal={IEEE Transactions on Pattern Analysis and Machine Intelligence},
  volume={41},
  number={8},
  pages={2008--2026},
  year={2019},
}

@misc{tran2016edward,
  title={Edward: A library for probabilistic modeling, inference, and criticism}, 
      author={Dustin Tran and Alp Kucukelbir and Adji B. Dieng and Maja Rudolph and Dawen Liang and David M. Blei},
      year={2017},
      eprint={1610.09787},
      archivePrefix={arXiv},
      primaryClass={stat.CO},
      url={https://arxiv.org/abs/1610.09787}, 
}

@article{blaiotta2016variational,
  title = {Variational Inference for Medical Image Segmentation},
  author = {Blaiotta, Claudia and Cardoso, M. Jorge and Ashburner, John},
  year = {2016},
  journal = {Computer Vision and Image Understanding},
  volume = {151},
  pages = {14--28},
}

@article{palma2025data,
  title = {Data-Driven Seasonal Climate Predictions via Variational Inference and Transformers},
  author = {Palma, Llu{\'i}s and Peraza, Alejandro and {Civantos-Prieto}, David and Duarte, Amanda and Materia, Stefano and Mu{\~n}oz, {\'A}ngel G. and {Pe{\~n}a-Izquierdo}, Jes{\'u}s and Romero, Laia and Soret, Albert and Donat, Markus G.},
  year = 2026,
  month = jan,
  journal = {npj Climate and Atmospheric Science},
  volume = {9},
  number = {1},
  pages = {48},
  issn = {2397-3722},
  doi = {10.1038/s41612-026-01320-z}
}

@article{kavian2024new,
  author={Kavianihamedani, Hossein and Quinn, Julianne D. and Smith, Jared D.},
  journal={IEEE Access}, 
  title={New Diagnostic Assessment of MCMC Algorithm Effectiveness, Efficiency, Reliability, and Controllability}, 
  year={2024},
  volume={12},
  pages={42385-42400},
}

@article{arian2020encoded,
journal={Mathematics and Computers in Simulation (MATCOM)},
author={Arian, Hamid and Moghimi, Mehrdad and Tabatabaei, Ehsan and Zamani, Shiva},
title={Encoded {{Value-at-Risk}}: {{A}} machine learning approach for portfolio risk measurement},
year={2022},
pages={500-525},
volume={202},
number={C},
}

@article{jaakkola1997variational,
  title = {A Variational Approach to {B}ayesian Logistic Regression Models and their Extensions},
  author = {Jaakkola, Tommi S. and Jordan, Michael I.},
  journal = {Proceedings of the Sixth International Workshop on Artificial Intelligence and Statistics},
  pages = {283--294},
  year = {1997},
  editor = {Madigan, David and Smyth, Padhraic},
  volume = {R1},
  series = {Proceedings of Machine Learning Research},
  month = {04--07 Jan},
  publisher = {PMLR},
  note = {Reissued by PMLR on 30 March 2021.}
}

@misc{thompson2010graphical,
  title={Graphical Comparison of MCMC Performance}, 
      author={Madeleine B. Thompson},
      year={2010},
      eprint={1011.4457},
      archivePrefix={arXiv},
      primaryClass={stat.CO},
            url={https://arxiv.org/abs/1011.4457}, 

}

@misc{vehtari_2022_advi_dropped,
  author       = {Vehtari, Aki},
  title        = {ADVI: Too many dropped evaluations even for well behaved models},
  year         = {2022},
  howpublished = {The Stan Forums},
  note         = {Accessed: 2025-04-23},
  url = {https://discourse.mc-stan.org/t/advi-too-many-dropped-evaluations-even-for-well-behaved-models/24338}
}

@misc{cmdstanpy_dev_2022_vi_example,
  author       = {{CmdStanPy Developers}},
  title        = {Variational Inference in Stan},
  year         = {2022},
  howpublished = {CmdStanPy Documentation v1.0.7},
  note         = {Accessed: 2026-05-27},
  url = {https://cmdstanpy.readthedocs.io/en/v1.0.7/users-guide/examples/Variational%20Inference.html}
}

@misc{stackexchange110369,
  title={Why and how Variational Inference underestimates variance?},
  author={Anonymous},
  howpublished={Data Science Stack Exchange},
  year={2023},
  note={Accessed: 2025-04-23},
  url={https://datascience.stackexchange.com/questions/110369/why-and-how-variational-inference-underestimates-variance}
}

@misc{mcstan32449,
  title={Parameters way off with variational inference, not sure why},
  author={Herrera, Daniel and Modr\'{a}k, Martin},
  howpublished={The Stan Forums},
  year={2023},
  note={Accessed: 2025-04-23},
  url = {https://discourse.mc-stan.org/t/parameters-way-off-with-variational-inference-not-sure-why/32449}
}

@inproceedings{salimans2015markovchainmontecarlo,
author = {Salimans, Tim and Kingma, Diederik P. and Welling, Max},
title = {Markov Chain {Monte Carlo} and variational inference: bridging the gap},
year = {2015},
booktitle = {Proceedings of the 32nd International Conference on Machine Learning},
pages = {1218–1226},
  editor = 	 {Bach, Francis and Blei, David},
  volume = 	 {37},
  address = 	 {Lille, France},
  month = 	 {07--09 Jul},
  publisher =    {PMLR},
}

@misc{mcstan4150,
  title={Question on variational inference},
  author={Ho, Bruce and Gabry, Jonah and Bonnevie, Rasmus and Carpenter, Bob},
  howpublished={The Stan Forums},
  year={2018},
  note={Accessed: 2025-04-23},
  url = {https://discourse.mc-stan.org/t/question-on-variational-inference/4150}
}

@misc{arvizissue498,
  title={Issue \#6271: Improve advice / diagnostics when VI gives terrible results},
  author={Ricardo Vieira},
  howpublished={GitHub},
  year={2022},
  note={Accessed: 2025-04-23},
  url={https://github.com/pymc-devs/pymc/issues/6271}
}

@book{bishop2007ch10,
  author = {Bishop, Christopher M.},
  title = {Pattern Recognition and Machine Learning},
  publisher = {Springer},
  series = {Information Science and Statistics},
  year = {2006},
  isbn = {0387310738}
}

@article{nemeth2019stochastic,
  title = {Stochastic Gradient Markov Chain Monte Carlo},
  author = {Nemeth, Christopher and Fearnhead, Paul},
  year = {2021},
  journal = {Journal of the American Statistical Association},
  volume = {116},
  number = {533},
  eprint = {https://doi.org/10.1080/01621459.2020.1847120},
  pages = {433--450},
  publisher = {ASA Website},
  
}

@article{Roberts2004general,
  title={General state space Markov chains and MCMC algorithms},
  volume={1},
  journal={Probability Surveys},
  pages = {20--71},
  author={Roberts, Gareth O. and Rosenthal, Jeffrey S.},
  year={2004}
}

@article{robbins1951stochastic,
  author  = {Herbert Robbins and Sutton Monro},
  title   = {A Stochastic Approximation Method},
  journal = {The Annals of Mathematical Statistics},
  volume  = {22},
  number  = {3},
  pages   = {400--407},
  year    = {1951},
  month   = sep,
  
}

@misc{cranlogs,
  author       = {{CRAN Logs}},
  title        = {{CRAN download counts: API for CRAN package download statistics}},
  howpublished = {\url{https://cranlogs.r-pkg.org/}},
  note         = {Accessed: 2025-10-30},
  year = {2025}
}

@misc{stan2025usersguide,
  title        = {Stan User's Guide},
  author       = {{Stan Development Team}},
  year         = {2025},
  note         = {Version 2.37.0},
  url  = {https://mc-stan.org}
}

@misc{bambi_docs_2025,
  title        = {Bambi: Bayesian Model-Building Interface},
  author       = {{Bambi Developers}},
  year         = {2025},
  howpublished = {\url{https://bambinos.github.io/bambi/}},
  note         = {Accessed: 2025-10-02}
}

@misc{pypistats_zenodo,
  title        = {pypistats},
  author       = {van Kemenade, Hugo},
  year         = {2024},
  howpublished    = {Zenodo},
  note = {doi: 10.5281/zenodo.10254758}
}

@misc{speagle2020conceptual,
      title={A Conceptual Introduction to {Markov} Chain {Monte Carlo} Methods}, 
      author={Joshua S. Speagle},
      year={2020},
      eprint={1909.12313},
      archivePrefix={arXiv},
      primaryClass={stat.OT},
      url={https://arxiv.org/abs/1909.12313}, 
}

@inproceedings{
phan2019composable,
title={{Composable Effects for Flexible and Accelerated Probabilistic Programming in NumPyro}},
author={Du Phan and Neeraj Pradhan and Martin Jankowiak},
booktitle={Program Transformations for ML Workshop at NeurIPS 2019},
year={2019}
}

@misc{dillon2017tensorflow,
title={TensorFlow Distributions}, 
      author={Joshua V. Dillon and Ian Langmore and Dustin Tran and Eugene Brevdo and Srinivas Vasudevan and Dave Moore and Brian Patton and Alex Alemi and Matt Hoffman and Rif A. Saurous},
      year={2017},
      eprint={1711.10604},
      archivePrefix={arXiv},
      primaryClass={cs.LG},
      url={https://arxiv.org/abs/1711.10604}
}

@inproceedings{abadi2016tensorflow,
  author = {Mart{\'\i}n Abadi and Paul Barham and Jianmin Chen and Zhifeng Chen and Andy Davis and Jeffrey Dean and Matthieu Devin and Sanjay Ghemawat and Geoffrey Irving and Michael Isard and Manjunath Kudlur and Josh Levenberg and Rajat Monga and Sherry Moore and Derek G. Murray and Benoit Steiner and Paul Tucker and Vijay Vasudevan and Pete Warden and Martin Wicke and Yuan Yu and Xiaoqiang Zheng},
title = {{TensorFlow}: A System for {Large-Scale} Machine Learning},
booktitle = {12th USENIX Symposium on Operating Systems Design and Implementation (OSDI 16)},
year = {2016},
isbn = {978-1-931971-33-1},
address = {Savannah, GA},
pages = {265--283},
url = {https://www.usenix.org/conference/osdi16/technical-sessions/presentation/abadi},
publisher = {USENIX Association},
month = nov
}

@misc{numpyro_elbo_2019,
  title        = {Source code for numpyro.infer.elbo},
  author       = {{Contributors to the Pyro Project}},
  howpublished = {\url{https://num.pyro.ai/en/0.4.0/_modules/numpyro/infer/elbo.html}},
  year         = {2019},
  version      = {0.4.0}
}

@misc{pymc_obj_n_mc_advi_2018,
  title        = {Default value used for obj\_n\_mc in ADVI},
  author       = {Lao, Junpeng},
  howpublished = {PyMC Discourse},
  year         = {2018},
  month        = {May},
  day          = {16},
  note         = {Accessed: 2025-04-23},
  url = {https://discourse.pymc.io/t/default-value-used-for-obj-n-mc-in-advi/1227}
}

@misc{pymc_repo_2025,
  title        = {{PyMC: Probabilistic Programming in Python}},
  author       = {{PyMC Developers}},
  howpublished = {\url{https://github.com/pymc-devs/pymc}},
  year         = {2025}
}

@misc{pymc_how_to_initialize_advi_2023,
  title        = {How to initialize ADVI},
  author       = {Christopher Fonnesbeck},
  howpublished = {PyMC Discourse},
  year         = {2023},
  month        = {August},
  day          = {8},
  note         = {Accessed: 2025-04-23},
  url = {https://discourse.pymc.io/t/how-to-initialize-advi/12693}
}

@InProceedings{wang2005inadequacy,
  title = 	 {Inadequacy of interval estimates corresponding to variational  Bayesian approximations},
  author =       {Wang, Bo and Titterington, D. M.},
  booktitle = 	 {Proceedings of the Tenth International Workshop on Artificial Intelligence and Statistics},
  pages = 	 {373--380},
  year = 	 {2005},
  editor = 	 {Cowell, Robert G. and Ghahramani, Zoubin},
  volume = 	 {R5},
  series = 	 {Proceedings of Machine Learning Research},
  month = 	 {06--08 Jan},
  publisher =    {PMLR}
}

@article{giordano2015linear,
  author    = {Giordano, Ryan and Broderick, Tamara
               and Jordan, Michael I.},
  title     = {Linear Response Methods for Accurate Covariance Estimates from Mean Field Variational Bayes},
  journal = {Advances in Neural Information Processing Systems},
  volume    = {28},
  pages     = {1441--1449},
  year      = {2015}
}

@article{jerfel2021variational,
  title = {Variational Refinement for Importance Sampling Using the Forward Kullback–Leibler Divergence},
  author = {Jerfel, Ghassen and Wang, Serena and Fannjiang, Clara and Heller, Katherine A. and Ma, Yian and Jordan, Michael I.},
  journal= {Proceedings of the Thirty-Seventh Conference on Uncertainty in Artificial Intelligence (UAI 2021)},
  pages = {1819--1829},
  year = {2021},
  publisher = {PMLR},
  volume = {161}
}

@book{mackay_2003_ita,
  author    = {MacKay, David J. C.},
  title     = {Information Theory, Inference and Learning Algorithms},
  publisher = {Cambridge University Press},
  year      = {2003},
}

@incollection{turner_sahani_2011_vem,
  author    = {Turner, Richard E. and Sahani, Maneesh},
  title     = {Two Problems with Variational Expectation Maximisation for Time-Series Models},
  booktitle = {Bayesian Time Series Models},
  editor    = {Barber, David and Cemgil, A. Taylan and Chiappa, Silvia},
  chapter   = {5},
  pages     = {109--130},
  publisher = {Cambridge University Press},
  year      = {2011},
}

@misc{numpyro_docs_stable_2025,
  title        = {{NumPyro Documentation}},
  author       = {{NumPyro Developers}},
  howpublished = {\url{https://num.pyro.ai/en/stable/}},
  year         = {2025}
}

@misc{pymc_transforms_api_2025,
  title        = {{Transformations — PyMC Distributions API}},
  author       = {{PyMC Development Team}},
  howpublished = {\url{https://www.pymc.io/projects/docs/en/latest/api/distributions/transforms.html}},
  year         = {2025}
}

@misc{numpyro_github_2025,
  title        = {{NumPyro: Probabilistic Programming with NumPy and JAX}},
  author       = {{NumPyro Developers}},
  howpublished = {\url{https://github.com/pyro-ppl/numpyro}},
  year         = {2025}
}

@misc{mcstan_cmdstanpy_variational_density_params_2024,
  title        = {CmdStanPy Variational - How to get density parameters?},
  author       = {madelynandersen},
  howpublished = {The Stan Forums},
  year         = {2024},
  month        = {November},
  day          = {17},
}

@article{duchi2011adaptive,
  author  = {John Duchi and Elad Hazan and Yoram Singer},
  title   = {Adaptive Subgradient Methods for Online Learning and Stochastic Optimization},
  journal = {Journal of Machine Learning Research},
  year    = {2011},
  volume  = {12},
  number  = {61},
  pages   = {2121--2159},
}

@article{kingma2015adam,
  title={Adam: A method for stochastic optimization},
  author={Kingma, Diederik P and Ba, Jimmy},
  journal={International Conference on Learning Representations (ICLR)},
  year={2015},
  note={arXiv:1412.6980}
}

@article{reddi2018convergence,
  title={On the convergence of Adam and beyond},
  author={Reddi, Sashank J and Kale, Satyen and Kumar, Sanjiv},
  journal={International Conference on Learning Representations (ICLR)},
  year={2018}
}

@article{wilson2017marginal,
  title={The marginal value of adaptive gradient methods in machine learning},
  author={Wilson, Ashia C and Roelofs, Rebecca and Stern, Mitchell and Srebro, Nathan and Recht, Benjamin},
  journal={Advances in Neural Information Processing Systems (NeurIPS)},
  volume={30},
  year={2017},
}

@article{Giordano2024BlackBox,
    author  = {Ryan Giordano and Martin Ingram and Tamara Broderick},
  title   = {Black Box Variational Inference with a Deterministic Objective: Faster, More Accurate, and Even More Black Box},
  journal = {Journal of Machine Learning Research},
  year    = {2024},
  volume  = {25},
  number  = {18},
  pages   = {1--39},
}

@article{Mason2018BayesianFramework,
  author    = {Alexina J. Mason and Manuel Gomes and Richard Grieve and James R. Carpenter},
  title     = {A Bayesian framework for health economic evaluation in studies with missing data},
  journal   = {Health Economics},
  year      = {2018},
  volume    = {27},
  number    = {11},
  pages     = {1670--1683},
  
  pmid      = {29969834},
  pmcid     = {PMC6220766}
}

@book{SASIntroBayes2013,
  author      = {{SAS Institute Inc.}},
  title   = {SAS/STAT\textregistered\, 13.1 User’s Guide: The MCMC Procedure},
  publisher   = {SAS Institute Inc.},
  address     = {Cary, NC},
  year        = {2013}
}

@book{RossBayesianReasoning2022,
  author    = {Kevin Davis Ross},
  title     = {Bayesian Reasoning and Methods},
  year      = {2022},
  publisher = {Open Book Publishers},
  address   = {Cambridge, UK}
}

@article{Davidai2012,
  author       = {Davidai, Shai and Gilovich, Thomas and Ross, Lee D.},
  title        = {The meaning of default options for potential organ donors},
  journal      = {Proceedings of the National Academy of Sciences of the United States of America},
  year         = {2012},
  volume       = {109},
  number       = {38},
  pages        = {15201--15205},
  
}

@article{Mehta2016,
  author       = {Mehta, Shivan J. and Troxel, Andrea B. and Marcus, Noora and Jameson, Christina and Taylor, Devon and Asch, David A. and Volpp, Kevin G.},
  title        = {Participation Rates With Opt-out Enrollment in a Remote Monitoring Intervention for Patients With Myocardial Infarction},
  journal      = {JAMA Cardiology},
  year         = {2016},
  volume       = {1},
  number       = {7},
  pages        = {847--848},
  
}

@incollection{Vaart_1998, 
title={Bayes Procedures}, 
booktitle={Asymptotic Statistics}, 
publisher={Cambridge University Press}, 
author={{van der Vaart}, A. W.}, 
year={1998}, 
pages={138–152}, 
}

@incollection{pushforward,
  title = {Sampling via Measure Transport: An Introduction},
  author = {Youssef Marzouk and Tarek Moselhy and Matthew Parno and Alessio Spantini},
  booktitle = {Handbook of Uncertainty Quantification},
  publisher = {Springer International Publishing},
  address = {Cham},
  year = {2016},
  pages = {1--41},
  editor = {R. Ghanem and D. Higdon and H. Owhadi},
  doi = {10.1007/978-3-319-11259-6_23-1}
}

@misc{keras_adam_optimizer_v3_3_3,
  author       = {{TensorFlow Developers}},
  title        = {{adam.py at v3.3.3 — Keras Optimizers (Adam) Source Code}},
  howpublished = {\url{https://github.com/keras-team/keras/blob/v3.3.3/keras/src/optimizers/adam.py}},
  year         = {2023},
  note         = {Accessed: 2026-01-02},
}

@misc{jax_example_libraries_optimizers_2026,
  author       = {{JAX Developers}},
  title        = {{example\_libraries/optimizers.py}},
  howpublished = {JAX Optimizers Example Code. \url{https://github.com/jax-ml/jax/blob/main/jax/example_libraries/optimizers.py}},
  year         = {2026},
  note         = {Accessed: 2026-01-02},
}

@misc{numpyro_svi_latest_2025,
  author       = {{NumPyro Developers}},
  title        = {{Stochastic Variational Inference (SVI)}},
  howpublished = {NumPyro Documentation.  \url{https://num.pyro.ai/en/latest/svi.html}},
  year         = {2025},
  note         = {Accessed: 2026-01-02},
}

@misc{tfp_probabilistic_pca_2024,
  author       = {{TFP Developers}},
  title        = {{Probabilistic PCA — TensorFlow Probability Example}},
  howpublished = {\url{https://www.tensorflow.org/probability/examples/Probabilistic_PCA}},
  year         = {2024},
  note         = {Accessed: 2026-01-02},
}

@misc{tfp_vi_fit_surrogate_posterior_2024,
  author       = {{TFP Developers}},
  title        = {tfp.vi.fit\_surrogate\_posterior — TensorFlow Probability API Documentation},
  howpublished = {\url{https://www.tensorflow.org/probability/api_docs/python/tfp/vi/fit_surrogate_posterior}},
  year         = {2024},
  note         = {Accessed: 2026-01-02},
}

@misc{tfp_release_notebook_0_12_1,
  author       = {{TFP Developers}},
  title        = {{TFP Release Notes notebook (0.12.1)}},
  howpublished = {\url{https://www.tensorflow.org/probability/examples/TFP_Release_Notebook_0_12_1}},
  year         = {2023},
  note         = {Accessed: 2026-01-02},
}

@misc{tfp_joint_distribution_2025,
  author       = {{TFP Developers}},
  title        = {Bayesian Modeling with Joint Distribution},
  howpublished = {\url{https://www.tensorflow.org/probability/examples/Modeling_with_JointDistribution}},
  year         = {2025},
  note         = {Accessed: 2026-01-02},
}

@misc{tfp_variational_inference_joint_dist_2024,
  author       = {{TFP Developers}},
  title        = {Variational Inference on Probabilistic Graphical Models with Joint Distributions},
  howpublished = {\url{https://www.tensorflow.org/probability/examples/Variational_Inference_and_Joint_Distributions}},
  year         = {2024},
  note         = {Accessed: 2026-01-02},
}

@misc{tfp_linear_mixed_effects_vi_2025,
  author       = {{TFP Developers}},
  title        = {Fitting Generalized Linear Mixed-Effects Models Using Variational Inference},
  howpublished = {\url{https://www.tensorflow.org/probability/examples/Linear_Mixed_Effects_Model_Variational_Inference}},
  year         = {2025},
  note         = {Accessed: 2026-01-02},
}

@article{LeCam1953,
  author  = {Le Cam, Lucien},
  title   = {On some asymptotic properties of maximum likelihood estimates and related Bayes' estimates},
  journal = {University of California Publications in Statistics},
  year    = {1953},
  volume  = {1},
  pages   = {277--329},
  mrnumber = {0054913}
}

@misc{numpyro_autoDAIS,
    author = {{NumPyro Developers}},
  title        = {{Example: AutoDAIS}},
  howpublished = {NumPyro Documentation.  \url{https://num.pyro.ai/en/stable/examples/dais_demo.html}},
  year         = {2025},
  note         = {Accessed: 2026-01-02},
}

@misc{pyro_svi_part_i,
    author = {{NumPyro Developers}},
  title        = {{SVI Part I: An Introduction to Stochastic Variational Inference in Pyro}},
  howpublished = {\url{https://pyro.ai/examples/svi_part_i.html}},
  year         = {2025},
  note         = {Accessed: 2026-01-02},
}

@misc{PyPA_PyPI_downloads,
  author       = {{Python Packaging Authority}},
  title        = {{Analyzing PyPI Package Downloads: Counting Package Downloads}},
  howpublished = {\url{https://packaging.python.org/en/latest/guides/analyzing-pypi-package-downloads/#counting-package-downloads}},
  note         = {Last updated November 12, 2025},
  year         = {2025}
}

@misc{campbell_placeholder,
    title={Large-scale empirical tuning and comparison of default optimizers for variational inference}, 
      author={Trevor Campbell and Jonathan H. Huggins and Kyurae Kim and Charles C. Margossian},
      year={2026},
      eprint={2606.07841},
      archivePrefix={arXiv},
      primaryClass={stat.CO},
}
\newpage
\appendix
\toplevel{Conjugate Benchmark Models}
\label{sec:benchmarks_conjugate}

This appendix collects analytically tractable conjugate models we may use as future benchmarks and details the conjugate beta--bernoulli model we use in our discussion of transformations and initialization. For each, we state the prior, likelihood, and closed-form posterior. These models support unit tests of VI implementations under default settings.

\secondlevel{Benchmark Models}
\thirdlevel{Beta--Bernoulli (Unknown Probability $\theta$)}
\label{sec:beta_bernoulli}

Consider binary observations $y_i \in \{0,1\}$ with a Bernoulli likelihood and a Beta prior on the success probability $\theta$:
\[
y_i \stackrel{\text{iid}}{\sim} \mathrm{Bernoulli}(\theta), 
\qquad
\theta \sim \mathrm{Beta}(\alpha_0, \beta_0),
\]
where $\mathrm{Beta}(\alpha,\beta)$ has density
$p(\theta) \propto \theta^{\alpha-1}(1-\theta)^{\beta-1}$ on $(0,1)$.
Let $S = \sum_{i=1}^n y_i$ be the number of observed successes. 
The posterior is
\[
\theta \mid y_{1:n} \sim \mathrm{Beta}\bigl(\alpha_0 + S,\; \beta_0 + n - S\bigr).
\]

The posterior mean and variance are
\[
\mu_n 
= \mathbb{E}[\theta \mid y_{1:n}]
= \frac{\alpha_0 + S}{\alpha_0 + \beta_0 + n}, 
\qquad
\mathrm{Var}(\theta \mid y_{1:n})
= \frac{(\alpha_0 + S)(\beta_0 + n - S)}{(\alpha_0 + \beta_0 + n)^2(\alpha_0 + \beta_0 + n + 1)}.
\]
For large $\alpha_0 + \beta_0 + n$, the Beta posterior is well approximated by a Gaussian distribution,
\[
\theta \mid y_{1:n} \;\approx\; 
\mathcal{N}\!\left(
  \frac{\alpha_0 + S}{\alpha_0 + \beta_0 + n},\;
  \frac{(\alpha_0 + S)(\beta_0 + n - S)}{(\alpha_0 + \beta_0 + n)^2(\alpha_0 + \beta_0 + n + 1)}
\right),
\]
with support truncated to $(0,1)$.
This benchmark stresses correct handling of $(0,1)$ support via log-odds or sigmoid transforms and sensitivity to skewed posteriors near the boundaries.

\secondlevel{Future Conjugate Benchmarks}
\thirdlevel{Gamma--Poisson (Unknown Rate $\lambda$)}
\label{sec:gamma_poisson}
With Poisson likelihood and Gamma prior on the rate,
\[
y_i \stackrel{\text{iid}}{\sim} \mathrm{Poisson}(\lambda),\qquad
\lambda \sim \mathrm{Gamma}(\alpha_0,\beta_0),
\]
using the rate parameterization $\mathrm{Gamma}(\alpha,\beta)$ with density
$p(\lambda)\propto \lambda^{\alpha-1}e^{-\beta\lambda}$, $\lambda>0$.
Let $S=\sum_{i=1}^n y_i$. The posterior is
\[
\lambda \mid y_{1:n} \sim \mathrm{Gamma}\bigl(\alpha_0+S,\;\beta_0+n\bigr).
\]
This is a natural check for correct support transforms to $(0,\infty)$ and stable variance estimation.

\thirdlevel{Gamma--Exponential (Unknown Rate $\lambda$)}
\label{sec:gamma_exponential}
With Exponential likelihood and Gamma prior on the rate,
\[
y_i \stackrel{\text{iid}}{\sim} \mathrm{Exponential}(\lambda),\qquad
\lambda \sim \mathrm{Gamma}(\alpha_0,\beta_0).
\]
Let $T=\sum_{i=1}^n y_i$. The posterior is
\[
\lambda \mid y_{1:n} \sim \mathrm{Gamma}\bigl(\alpha_0+n,\;\beta_0+T\bigr).
\]
This benchmark stresses correctness of the positive-real bijector and sensitivity to skew.

\thirdlevel{Inverse-Gamma--Normal (Unknown Variance $\sigma^2$, Known Mean $\mu$)}
\label{sec:inv_gamma_normal}
Assume the mean $\mu$ is known and the variance is unknown:
\[
y_i \stackrel{\text{iid}}{\sim} \mathcal{N}(\mu,\sigma^2),\qquad
\sigma^2 \sim \mathrm{Inv\text{-}Gamma}(\alpha_0,\beta_0),
\]
with $\mathrm{Inv\text{-}Gamma}(\alpha,\beta)$ density $p(s^2)\propto (s^2)^{-(\alpha+1)} e^{-\beta/s^2}$ for $s^2>0$.
Let $S=\sum_{i=1}^n (y_i-\mu)^2$. The posterior is
\[
\sigma^2 \mid y_{1:n} \sim \mathrm{Inv\text{-}Gamma}\!\left(\alpha_0+\frac{n}{2},\;\beta_0+\frac{S}{2}\right).
\]
This is a test of variance-parameter transforms (log vs.\ softplus) and gradient noise.

\thirdlevel{High-Dimensional Gaussian--Gaussian (Unknown Mean, Known Covariance)}
\label{sec:highdim_mvn}
Let $y_i \in \mathbb{R}^d$ and $\Sigma$ be known, positive definite. The model is
\[
y_i \stackrel{\text{iid}}{\sim} \mathcal{N}_d(\mu,\Sigma),\qquad
\mu \sim \mathcal{N}_d(\mu_0,\Sigma_0).
\]
Define $\bar{y}=\frac{1}{n}\sum_{i=1}^n y_i$. The posterior is
\[
\Sigma_p = \bigl(n\Sigma^{-1} + \Sigma_0^{-1}\bigr)^{-1},\qquad
\mu_p = \Sigma_p\bigl( n\Sigma^{-1}\bar{y} + \Sigma_0^{-1}\mu_0 \bigr),
\]
so that
\[
\mu \mid y_{1:n} \sim \mathcal{N}_d(\mu_p,\Sigma_p).
\]
This benchmark stresses scaling with dimension: mean-field guides ignore posterior correlations,
while full-rank guides can capture them but require $O(d^2)$ parameters.

\FloatBarrier

\toplevel{Package Settings}
\label{sec:defaults}

\begin{table}[htbp]
\footnotesize
\setlength{\tabcolsep}{4pt}
\centering
\caption{PyMC Transformation Mappings}
\label{fig:PyMC_transformation_table}
\begin{tabularx}{\linewidth}{@{}Y T Y@{}}
\toprule
\textbf{Constraint / Case} & \textbf{Transformation} & \textbf{Notes} \\
\midrule
\texttt{PositiveContinuous}
  & \texttt{transforms.log}
  & $\mathbb{R}\!\to\!(0,\infty)$ (log) \\
\addlinespace
\texttt{UnitContinuous}
  & \texttt{transforms.logodds}
  & $\mathbb{R}\!\to\!(0,1)$ (log-odds) \\
\addlinespace
\texttt{CircularContinuous}
  & \texttt{transforms.circular}
  & Wraps to $[0,2\pi)$ \\
\addlinespace
\texttt{BoundedContinuous}
  & \texttt{transforms.Interval}
  & $\mathbb{R}\!\to\!(a,b)$ \\
\midrule
\multicolumn{3}{@{}c@{}}{\textbf{Special Cases}} \\
\midrule
Scale parameters (VI)
  & \texttt{softplus (rho2sigma)}
  & Ensures positive scale \\
\addlinespace
Ordered vectors
  & \texttt{OrderedTransform()}
  & Cumulative-sum ordering \\
\addlinespace
Simplex
  & \texttt{SumTo1}
  & Sum-to-1 projection \\
\addlinespace
Cholesky covariance
  & \texttt{CholeskyCovPacked}
  & Log on diagonal of Cholesky factor \\
\bottomrule
\end{tabularx}
\end{table}

\begin{table}[htbp]
\footnotesize
\setlength{\tabcolsep}{4pt}
\centering
\caption{NumPyro Transformation mappings}
\label{fig:NumPyro_transformation_table}
\begin{tabularx}{\linewidth}{@{}Y T Y@{}}
\toprule
\textbf{Constraint (constraints.)} & \textbf{Transformation} & \textbf{Notes / Examples} \\
\midrule
\texttt{positive} & \texttt{ExpTransform()} & Exponential, Gamma \\
\texttt{nonnegative} & \texttt{ExpTransform()} & Same as positive \\
\texttt{greater\_than(lower\_bound)} &
\texttt{ComposeTransform([ ExpTransform(), AffineTransform(lower\_bound, 1) ])} &
Shifted; ensures values $>$ lower bound \\
\texttt{less\_than(upper\_bound)} &
\texttt{ComposeTransform([ ExpTransform(), AffineTransform(upper\_bound, -1) ])} &
Ensures values $<$ upper bound \\
\texttt{unit\_interval} & \texttt{SigmoidTransform()} & Maps $\mathbb{R}\!\to\!(0,1)$ \\
\texttt{interval(lower, upper)} &
\texttt{ComposeTransform([ SigmoidTransform(), AffineTransform(lower, scale) ])} &
Uniform$(a,b)$ \\
\texttt{simplex} & \texttt{StickBreakingTransform()} & Simplex (e.g.\, Dirichlet) \\
\texttt{ordered\_vector} & \texttt{OrderedTransform()} & Enforces increasing order \\
\texttt{positive\_definite} &
\texttt{ComposeTransform([ LowerCholeskyTransform(), CholeskyTransform().inv ])} &
Positive-definite matrices \\
\texttt{real} & \texttt{IdentityTransform()} & Normal (no transform) \\
\midrule
\multicolumn{3}{@{}c@{}}{\textbf{Distribution Examples}} \\
\midrule
Normal & \texttt{constraints.real} & No transform \\
Exponential & \texttt{constraints.positive} & \texttt{ExpTransform()} \\
Gamma & \texttt{constraints.positive} & \texttt{ExpTransform()} \\
Uniform$(a,b)$ &
\texttt{ComposeTransform([ SigmoidTransform(), AffineTransform(lower, scale) ])} & — \\
\bottomrule
\end{tabularx}
\end{table}

\FloatBarrier

\toplevel{Numerical Stability and Transformations}\label{sec:numerical-stability}

This appendix records mathematical details relevant to variance parameterizations in
variational inference and summarizes standard transformations used to map between
unconstrained and constrained parameter spaces. We focus on functional forms,
derivatives, and log-Jacobians that appear in ELBO calculations.

\secondlevel{Variance Parameterizations}

For a univariate latent variable $z$, we take the variational family to be
\[
  q_{\phi}(z) = \mathcal{N}(z \mid m, \sigma^{2}), 
  \qquad 
  \phi = (m, \omega),
\]
with $\omega$ an unconstrained scale parameter. Two parameterizations are commonly used:
\[
  \sigma = \exp(\omega),
  \qquad
  \sigma = \operatorname{softplus}(\omega)
  = \log(1+e^{\omega}).
\]

Under the reparameterization
\[
  z = m + \sigma\,\epsilon,
  \qquad \epsilon \sim \mathcal{N}(0,1),
\]
the ELBO is
\[
  \mathcal{L}(\phi)
  = \mathbb{E}_{\epsilon}[\log p(y,z)]
    - \mathbb{E}_{\epsilon}[\log q_{\phi}(z)].
\]

\thirdlevel{Log-Scale Parameterization}

With $\sigma=\exp(\omega)$,
\[
  \frac{\partial \sigma}{\partial \omega} = \exp(\omega),
  \qquad
  \frac{\partial}{\partial\omega}(m+\sigma\epsilon)
  = \sigma\epsilon.
\]

The entropy term
\[
  -\log q_{\phi}(z)
  = \tfrac12\Bigl(\frac{z-m}{\sigma}\Bigr)^{2}
    + \log\sigma + \tfrac12\log(2\pi)
\]
satisfies
\[
  \frac{\partial}{\partial\omega}[-\log q_{\phi}(z)] = 1.
\]

Thus,
\[
  \nabla_{\omega}\mathcal{L}
  = \mathbb{E}_{\epsilon}[
      \partial_{z}\log p(y,z)\,\sigma\epsilon
    ] + 1.
\]

\paragraph{Usage Notes.}
The log-scale parameterization has an unbounded derivative and therefore amplifies
large values of $\epsilon$. 

\thirdlevel{Softplus Parameterization}

For $\sigma=\operatorname{softplus}(\omega)$,
\[
  \frac{\partial \sigma}{\partial \omega}
  = \operatorname{sigmoid}(\omega),
\qquad
  \frac{\partial}{\partial\omega}(m+\sigma\epsilon)
  = \epsilon\,\operatorname{sigmoid}(\omega).
\]

For the entropy term,
\[
  \frac{\partial}{\partial\omega}[-\log q_{\phi}(z)]
  = \frac{\operatorname{sigmoid}(\omega)}{\operatorname{softplus}(\omega)}.
\]

The ELBO gradient becomes
\[
  \nabla_{\omega}\mathcal{L}
  =
  \mathbb{E}_{\epsilon}\!\bigl[
     \partial_{z}\log p(y,z)\,\epsilon\,\operatorname{sigmoid}(\omega)
  \bigr]
  +
  \frac{\operatorname{sigmoid}(\omega)}{\operatorname{softplus}(\omega)}.
\]

\paragraph{Usage Notes.}
The bounded derivative $\operatorname{sigmoid}(\omega)\in(0,1)$ moderates the effect
of large $\epsilon$ and smooths behavior near both small and large $\sigma$. This
parameterization is widely employed in modern VI implementations for scale variables.

\secondlevel{Transformations for Constrained Parameters}\label{sec:softplus}

Many probabilistic models require transforming an unconstrained variable
$u\in\mathbb{R}$ to a constrained domain via $x=f(u)$. The change of variables is
\[
  p_U(u)
  = p_X(f(u))\left|\frac{dx}{du}\right|,
  \qquad
  \log\left|\frac{dx}{du}\right|.
\]

We record formulas for standard transformations.

\thirdlevel{Exponential Transform}

\[
  x = \exp(u),
  \qquad u \in \mathbb{R}, \; x \in (0,\infty).
\]

\[
  f^{-1}(x) = \log x,
  \qquad
  \frac{dx}{du} = x,
  \qquad
  \log\left|\frac{dx}{du}\right| = u.
\]

\paragraph{Usage Notes.}
This mapping enforces positivity and appears in reparameterizations of parameters including the rate variable of poisson or exponential distributions.

\thirdlevel{Logarithm Transform}

\[
  u=\log x,
  \qquad x \in (0,\infty).
\]

\[
  f^{-1}(u)=\exp(u),
  \qquad
  \frac{du}{dx} = \tfrac{1}{x},
  \qquad
  \log\left|\frac{du}{dx}\right| = -\log x.
\]

\paragraph{Usage Notes.}
The transform is the inverse of the exponential mapping. It is sometimes used to avoid directly evaluating very large or very small values that approach numerical overflow or underflow, respectively.

\thirdlevel{Sigmoid Transform}

\[
  x = \sigma(u)=\frac{1}{1+e^{-u}},
  \qquad u\in\mathbb{R},\; x\in(0,1).
\]

\[
  f^{-1}(x) = \log\!\frac{x}{1-x},
  \qquad
  \frac{dx}{du} = x(1-x),
  \qquad
  \log\left|\frac{dx}{du}\right| = \log x + \log(1-x).
\]

\paragraph{Usage Notes.}
The sigmoid transform is the standard way to parameterize probabilities.
Its derivative vanishes near boundaries, which compresses extreme regions of
the constrained space into broad regions of $u$.

\thirdlevel{Log-Odds (Logit) Transform}

\[
  u = \log\!\frac{x}{1-x},
  \qquad x\in(0,1).
\]

\[
  f^{-1}(u)=\sigma(u),
  \qquad
  \frac{du}{dx} = \frac{1}{x(1-x)},
  \qquad
  \log\left|\frac{du}{dx}\right| = -\log x - \log(1-x).
\]

\paragraph{Usage Notes.}
The logit transform is the inverse of the sigmoid mapping. It linearizes deviations
around interior probabilities but expands neighborhoods of $x\to0$ or $x\to1$.

\thirdlevel{Softplus Transform}

\[
  x = \operatorname{softplus}(u)=\log(1+e^{u}),
  \qquad u\in\mathbb{R},\; x\in(0,\infty).
\]

\[
  f^{-1}(x)=\log(e^{x}-1),
  \qquad
  \frac{dx}{du}=\operatorname{sigmoid}(u),
  \qquad
  \log\left|\frac{dx}{du}\right|
  = -\log\!\bigl(1+e^{-u}\bigr).
\]

\paragraph{Usage Notes.}
Softplus is a smooth alternative to $\exp(u)$ with slower growth for large $u$ and
a positive lower envelope for small $u$. It is used extensively for stabilizing
scale-parameter optimization.

\thirdlevel{Interval Transform}

For any finite interval $(a,b)$,
\[
  x = a + (b-a)\,\sigma(u),
\]
\[
  \log\left|\frac{dx}{du}\right|
  = \log(b-a) + \log\sigma(u) + \log(1-\sigma(u)).
\]

\paragraph{Usage Notes.}
This maps $\mathbb{R}$ to $(a,b)$ via an affine transformation of the sigmoid.
It is common in models with bounded parameters.

\thirdlevel{Higher-Dimensional Constructions}

Matrix-valued constraints frequently rely on elementwise transforms applied to
Cholesky factors. For example, diagonal entries may be parameterized by
$d_i = \operatorname{softplus}(u_i)$, with the total log-Jacobian equal to the sum
over coordinates.

\begin{table}[h]
\centering
\caption{Common transformations used for constrained parameters.}
\label{tab:summary-transforms}
\begin{tabular}{lccc}
\toprule
\textbf{Transform} 
& \textbf{Domain $\to$ Codomain} 
& \textbf{Inverse $f^{-1}(x)$} 
& $\displaystyle \log\!\left|\frac{dx}{du}\right|$ \\
\midrule
Exponential 
& $\mathbb{R} \to (0,\infty)$ 
& $\log x$ 
& $u$ \\
Logarithm 
& $(0,\infty) \to \mathbb{R}$ 
& $e^{u}$ 
& $-\log x$ \\
Sigmoid 
& $\mathbb{R} \to (0,1)$ 
& $\log\!\tfrac{x}{1-x}$ 
& $\log x + \log(1-x)$ \\
Log-odds 
& $(0,1) \to \mathbb{R}$ 
& $\tfrac{1}{1+e^{-u}}$ 
& $-\log x - \log(1-x)$ \\
Softplus 
& $\mathbb{R} \to (0,\infty)$ 
& $\log(e^{x}-1)$ 
& $-\log(1+e^{-u})$ \\
\bottomrule
\end{tabular}
\end{table}

\FloatBarrier


\end{document}